%% file: main.tex
\documentclass[acmsmall]{acmart}

\setcopyright{acmcopyright}
\copyrightyear{xxxx}
\acmYear{2022}
\acmDOI{XXXXXXX.XXXXXXX}

\acmJournal{CSUR}
\acmVolume{xx}
\acmNumber{x} 
\acmArticle{111}
\acmMonth{12}

\usepackage{enumitem}
\newlist{rqs}{enumerate}{1}
\setlist[rqs]{label*=\textbf{RQ\arabic*~}}

\usepackage{csquotes}

\usepackage{hyperref}

\usepackage{booktabs,tabularx,makecell}

\usepackage{array}

\usepackage{graphicx}
\graphicspath{{images/}}

\usepackage{float}

\usepackage[labeled, resetlabels]{multibib}
\newcites{S}{Primary Studies}
\usepackage{tikz}

\begin{document}

\definecolor{colorA}{RGB}{114,141,196}
\definecolor{colorB}{RGB}{214,104,99}
\definecolor{colorC}{RGB}{109,179,147}
\definecolor{colorD}{RGB}{199,123,227}
\definecolor{colorE}{RGB}{255,193,116}
\definecolor{colorF}{RGB}{179,140,121}

%
% -------------------------------------------------------------------------- %
% Title, authors, abstract, keywords
% -------------------------------------------------------------------------- %
%

\title{40 Years of Designing Code Comprehension Experiments: A Systematic Mapping Study}

\author{Marvin Wyrich}
\email{wyrich@cs.uni-saarland.de}
\orcid{0000-0001-8506-3294}
\author{Justus Bogner}
\email{j.bogner@vu.nl}
\orcid{0000-0001-5788-0991}
\author{Stefan Wagner}
\email{stefan.wagner@iste.uni-stuttgart.de}
\orcid{0000-0002-5256-8429}
\affiliation{%
  \institution{University of Stuttgart}
  \streetaddress{Universitätsstraße 38}
  \city{Stuttgart}
  \country{Germany}
  \postcode{70569}
}

\renewcommand{\shortauthors}{Wyrich et al.}

\begin{abstract}
The relevance of code comprehension in a developer's daily work was recognized more than 40 years ago.
Consequently, many experiments were conducted to find out how developers could be supported during code comprehension and which code characteristics contribute to better comprehension.
Today, such studies are more common than ever.
While this is great for advancing the field, the number of publications makes it difficult to keep an overview.
Additionally, designing rigorous code comprehension experiments with human participants is a challenging task, and the multitude of design options can make it difficult for researchers, especially newcomers to the field, to select a suitable design.

We therefore conducted a systematic mapping study of 95 source code comprehension experiments published between 1979 and 2019.
By structuring the design characteristics of code comprehension studies, we provide a basis for subsequent discussion of the huge diversity of design options in the face of a lack of basic research on their consequences and comparability.
We describe what topics have been studied, as well as how these studies have been designed, conducted, and reported.
Frequently chosen design options and deficiencies are pointed out to support researchers of all levels of domain expertise in designing their own studies.
\end{abstract}




\maketitle

% silently cite all primary studies for the 2nd bibliography
\nociteS{*}

%
% -------------------------------------------------------------------------- %
% Introduction
% -------------------------------------------------------------------------- %
%

\section{Introduction}\label{sec:introduction}

Developers spend a lot of time understanding source code, i.e. finding meaning behind individual lines of code, to be able to work with it. Estimates of the average working time invested in source code comprehension range from 30 to 70\%~\cite{Minelli:2015:LastSummer,Xia:2018:Measuring}. Accordingly, there is great motivation among researchers to optimize this process through scientific research.
This effort can be seen in the establishment of the International Conference on Program Comprehension (ICPC),\footnote{\url{https://www.program-comprehension.org/}} which celebrated its 30\textsuperscript{th} anniversary in 2022. 

While some researchers investigate how source code can be written more comprehensibly, others analyze which contextual factors and individual characteristics influence code comprehension.
However, this is not as simple as it may sound: designing code comprehension studies with human participants is hard.
Even though some recommendations exist, they are largely a collection of arguments and considerations based on personal experience~\cite{Feitelson:2021:Considerations} or refer on a more abstract level to guidelines for general research methods like controlled experiments~\cite{Wohlin2012,Juristo2001}.

As a result, every researcher currently designs, conducts, and reports their code comprehension study quite differently.
For example, the differences in design begin with the concrete tasks given to the participants in a study, i.e., whether they only need to read code, fix a bug, or even extend the code.
Differences continue with the way in which the success of code comprehension is quantified, e.g., via the time required to solve a task, correctness in comprehension questions, subjective self-assessments by the participants, or psycho-physiological measurements~\cite{Oliveira:2020:Evaluating}.

We consider a certain diversity in study designs to be essential in good scientific practice, as complex research questions should be approached from different angles.
Currently, however, this leads to two major issues: first, it creates uncertainty when designing a new study, as it is not clear from the multitude of code comprehension studies what the majority of the program comprehension community currently agrees on as valid study designs~\cite{Siegmund:2016:PastPresentFuture,Siegmund:2015:Views}.
This makes it especially difficult for novice researchers to get familiar with this field.
Second, different study results are difficult to compare, let alone to incorporate into meta-analyses, in which study results are to be synthesized~\cite{Wohlin:2014:writing,Kitchenham2020}.

To address these issues, we present a systematic mapping study (SMS) which reveals the state-of-the-art in the design of code comprehension studies, and thus serves as a proxy for what the community currently considers acceptable study designs.
This analysis helps to reduce uncertainty if researchers simply follow the major tendencies among existing peer-reviewed studies on a certain topic when designing a code comprehension experiment.
Furthermore, our SMS discusses shortcomings in existing study designs and suggests ways to improve the state-of-the-art in the design and reporting of code comprehension studies.
This should also make it easier to compare different studies in the future.
The \textbf{contributions} of our work are therefore as follows:

\begin{itemize}
    \item A synthesis of differences and similarities in more than 15 study design characteristics of bottom-up code comprehension experiments, including an analysis of the research topics studied and the threats to validity discussed.
    \item A discussion of recurring issues in the design and reporting of code comprehension experiments, focusing on common design difficulties and those circumstances that make it hard to compare different studies. This discussion results in concrete proposals for improving the design and reporting of code comprehension experiments with human participants.
    \item A publicly available data set with a comprehensive list of 95 studies published between 1979 and 2019 that measured bottom-up code comprehension with human participants. The data set includes extracted design characteristics for each study.
\end{itemize}

With a vast landscape of information ahead, here is a quick roadmap to guide you through the paper. Section~\ref{sec:background} discusses what we mean by source code comprehension, describes a concrete example of a code comprehension experiment, and presents related work. Section~\ref{sec:methodology} explains our methodological approach to finding and analyzing code comprehension experiments and their design characteristics. We also introduce our specific research questions in this section. Section~\ref{sec:results} presents a detailed analysis of the design characteristics. In Section~\ref{sec:discussion}, we invite the reader to a discussion of the results before concluding in Section~\ref{sec:conclusion} with five action items that we should address as a code comprehension research community moving forward.

%
% -------------------------------------------------------------------------- %
% Background and Related Work
% -------------------------------------------------------------------------- %
%

\section{Background and Related Work}
\label{sec:background}
In this section, we describe necessary foundations regarding code comprehension research, and point out related work on the design of code comprehension experiments.

\subsection{Source Code Comprehension}
\label{sec:comprehension}

The cognitive process of code comprehension is still fairly unexplored.
Researchers in the field have a rough idea of what is meant when talking about comprehensible code or performance in code understanding.
However, a definition of the construct, let alone a comprehensive cognitive model, has not yet been established to the extent that primary studies would rely on it in their design.
As a result, what is measured is often implicitly defined by how it is measured, e.g., by the number of correctly answered questions about a C function or the time required to find a bug.

Nonetheless, there are starting points on which we can build to define the scope of this work.
For example, ISO/IEC 25010 describes a quality model for defining individual quality characteristics of a software product.
The quality attribute most related to code comprehensibility in this model is perhaps \textit{analyzability}, which is defined as: \enquote{Degree of effectiveness and efficiency with which it is possible to assess the impact on a product or system of an intended change to one or more of its parts, or to diagnose a product for deficiencies or causes of failures, or to identify parts to be modified}~\cite{25010:2011}.
This definition has some overlap with frequently used participant tasks in code comprehension studies.
However, it remains at a level of granularity that does not explicitly address that source code may need to be understood for all described intentions and what it means to have understood source code.

\citet{Boehm:1976:Quantitative} provide a more concrete definition of understandability in this regard in their early work on quantitative evaluation of software quality: \enquote{code possesses the property of understandability to the extent that its purpose is clear to the reviewer. This implies that variable names or symbols are used consistently, code modules are self-describing, control structure is simple or conforms to a prescribed standard, etc}.
The first sentence of \citeauthor{Boehm:1976:Quantitative}'s definition is the one that will guide us in this work.
The understandability of source code can be measured by how clear the purpose of the code is to the reader.
The second part of the definition provides some examples of code characteristics that would help improve comprehensibility.
These are common hypotheses that are the subject of the primary studies we will consider in our mapping study.

In this work, we furthermore limit ourselves to the investigation of studies in which participants had to understand a set of concrete lines of code to infer the intentions behind the code at a higher level of abstraction. 
This process is usually referred to as \textit{bottom-up comprehension}~\cite{Pennington:1987:Comprehension,OBrien:2004:BottomUp}.
The counterpart to this is called \textit{top-down comprehension}, a process in which developers refine hypotheses based on domain knowledge and documentation until the knowledge can be mapped to concrete code segments~\cite{Shaft:1995:Relevance}.
Using the works of \citet{Brooks:1983:TowardsTheory} and \citet{Soloway:1984:Empirical}, \citet{OBrien:2004:BottomUp} further distinguish top-down comprehension into expectation-based and inference-based comprehension, where the difference can be roughly summarized as whether beacons~\cite{Rist:1986:Plans,Wiedenbeck:1986:Beacons} in the code confirm or trigger a developer's (pre-generated) hypotheses.
For a more in-depth discussion, including these subtle differences, we refer to the work of~\citet{OBrien:2004:BottomUp}.

We focus on one type of comprehension process, which is bottom-up code comprehension, because there is enough reason to assume that different approaches are based on different cognitive models and, accordingly, study designs for different cognitive models are not necessarily comparable in nature.
In Section~\ref{subsec:incluexclu}, we present more details about our criteria for including papers in our mapping study, and already in Section~\ref{sec:exampleStudy} below, we show a concrete example of a code comprehension experiment that would be included.

\subsubsection*{Related Terms}
Now that we have a rough definition of comprehension, we explain how we use and distinguish certain related terms in the remainder of the paper.
First, we consider code comprehension studies as a subset of program comprehension studies.
If we look at the plethora of topics published at the International Conference on Program Comprehension, it becomes clear that researchers are not only concerned with the comprehensibility of source code, but also, e.g., with the comprehensibility of software architectures, documentation, and diagrams.

Second, like some of our peers in the field of program comprehension as well as text comprehension, we consider understandability and comprehensibility as synonyms, i.e., choosing one over the other is \enquote{purely a matter of linguistic variation}~\cite{Kintsch:1998:Comprehension}.

Third, comprehensibility in this context is a property of source code.
Comprehension, however, is the act or performance of making sense of code.
Studies that intend to provide insights into the influence of a treatment on the comprehensibility of code usually first measure comprehension, i.e., how well a participant understood source code under certain conditions.
The data collected can then be used to draw conclusions about the comprehensibility of the code.

Finally, comprehensibility, readability, and legibility are different constructs, and the program comprehension community is starting to distinguish these more explicitly.\footnote{Conscious distinction can be seen, e.g., in the call for this year's EMSE special issue on \enquote{code legibility, readability, and understandability} at \url{https://tinyurl.com/EMSECfP}.}
Code can be legible and readable due to being presented in a certain way and using familiar keywords.
However, readable code does not necessarily have to be understood by the reader.
In both code and text comprehension, there is evidence that comprehensibility and readability do not necessarily correlate~\cite{Borstler:2016:RoleMethodChains,Smith:1992:ReadabilityUnderstandability}.
In our mapping study, we restrict ourselves to studies in which code had to be understood.

\subsection{A Concrete Example of a Code Comprehension Experiment}
\label{sec:exampleStudy}

Let us take this opportunity to get a better sense of what a code comprehension experiment with human participants can look like, i.e., the kind of study that is interesting for our mapping study.
For this purpose, we exemplarily review the paper by \citet{Hofmeister:2017:Shorter}, who investigated the research question how identifier name length and semantics affect code comprehension (spoiler: words as identifier names are superior to letters and abbreviations in terms of comprehension speed).

After the authors motivated the problem, they needed to make a whole series of design decisions. First, how should the independent and dependent variables be measured, i.e., what is the operationalization of code comprehension? In this example, code comprehension performance was measured as the time required to successfully identify a defect in a presented code snippet. The identifier naming style was either \enquote*{Word}, \enquote*{Abbreviation}, or \enquote*{Letter}. Furthermore, each participant saw 6 snippets from a pool of 18 snippet variants. The authors created the snippets themselves to ensure that they could ultimately answer the research question. Their preferred experiment participants were professional C\# developers, who were recruited via Twitter, Xing, and technology industry conferences. Developers participated in the experiment via an online platform.

At this point, we covered only about a third of the key design decisions \citet{Hofmeister:2017:Shorter} discussed in their paper. All these decisions, e.g., how to measure comprehension, participant sampling strategy, comprehension task design, code snippet selection, remote experiment conduct, etc., could have been made differently. We do not know how easy it was for the authors to make these decisions. But we can infer from the detailed rationales for individual design decisions that these decisions were not made lightly following their gut instinct. 

It will become apparent in our systematic mapping study that different researchers make such decisions differently.
Since we are not the first to address differences in design decisions of code comprehension studies, we summarize what we build on in the following Section~\ref{sec:designingccstudies}.

\subsection{Designing Code Comprehension Studies}
\label{sec:designingccstudies}

In 2007, \citet{Di:2007:DesigningNext} proposed a working session to collaboratively launch empirical program comprehension studies.
Their motivation was that the design of such studies was complex due to the \enquote{large space of decisions} and that the community had to discuss what the best practices were.

In recent years, the debate about the design of program comprehension studies, which had subsided meanwhile, gained momentum again.
In 2016, \citet{Siegmund:2016:PastPresentFuture} suggested, based on a survey of program committee members at major software engineering venues, that the reason for the few new program comprehension studies may still be that \enquote{empirical evaluations of program comprehension (and human factors in general) are difficult to conduct}.

In 2021, \citet{Feitelson:2021:Considerations} published a series of \enquote{considerations and pitfalls in controlled experiments on code comprehension}.
For different design decisions like selecting code to be comprehended or the concrete comprehension tasks, the paper describes how these decisions can affect the results of a study, and exemplarily points to primary studies that implemented specific design decisions.
We believe that \citeauthor{Feitelson:2021:Considerations}'s work~\cite{Feitelson:2021:Considerations} is a valuable pragmatic approach for guiding researchers in the design of code comprehension experiments by compiling intuitive knowledge.
Our work will provide the necessary empirical data on how prevalent certain design decisions, and thus certain pitfalls, actually are.

Some publications on the analysis of existing program comprehension studies already exist (find a summary in Table~\ref{tab:rel_work}). 
\citet{Oliveira:2020:Evaluating} model program comprehension as a learning activity and map typical tasks of empirical human-centric studies on readability and legibility to an adaptation of Bloom's taxonomy of educational objectives.
Their goal is to identify the cognitive skills that are most frequently tested in readability and legibility studies.
We are particularly interested in their classification of commonly used tasks and measures, as we use it as a basis for our own classification of comprehension tasks.
Their data show that participants in readability and legibility studies most often have to provide information about the code, and correctness is favored as a response variable.

\citet{Schroter:2017:Comprehending} conducted a literature review of 540 ICPC papers published between 2006 and 2016.
They addressed three questions, namely what the primary studies examined in the context of program comprehension, what terminology they used to describe the evaluation (e.g., empirical study, user study, etc.), and whether they reported threats to validity.
Their investigation revealed that \textit{source code} and \textit{program behavior} are the most frequently addressed parts of program comprehension studies at ICPC.
Moreover, researchers would use \enquote{a diverse and often ambiguous terminology to report their evaluation}.
Finally, they found that the number of empirical studies reporting threats to validity has increased in recent years.
This study complements our work very well, since it deals with program comprehension studies in a broader scope and conducts similar investigations as we do.
Yet, we analyze the topic more deeply and consider numerous other study design features, as -- according to their comment on potential future work -- \citet{Schroter:2017:Comprehending} themselves intended \enquote{to analyze in more detail how evaluations on software comprehension are performed, for example, regarding typical evaluation tasks and comprehension measurement}.

Related to the analysis of threats to validity is a literature survey by \citet{Siegmund:2015:Confounding} that compiles a catalog of confounding parameters discussed in program comprehension studies.
Their survey includes papers from 13 journals and conferences over a period of 10 years and results in 39 confounding parameters.
In our systematic mapping study, we will use the resulting categories of confounders as a basis to categorize discussed threats to validity in code comprehension studies.
We will come back to this in more detail in Section~\ref{reported_limitations}.

\begin{table}[ht]
\caption{Overview of related survey work in the program comprehension field}
\label{tab:rel_work}
\begin{tabularx}{\columnwidth}{l X X}
\toprule
Secondary study & What they did & What we do in comparison \\
\midrule
\citet{Oliveira:2020:Evaluating} & Study of 54 comprehensibility, readability, and legibility papers. Focus on task and response variables. Subsequent mapping to Bloom's learning taxonomy. & Study of more design aspects. Strict focus on studies in which code had to be understood (not, for example, only read). However, we build on their task and measure classification.\\
\citet{Schroter:2017:Comprehending} & Study of 540 ICPC papers (2006--2016). Focus on terminology of evaluation, threats to validity. & Study of more design aspects over a larger time period (1979--2019). No restriction to a single venue. Instead, restriction to those studies where code snippets need to be understood.\\
\makecell{Siegmund and\\Schumann~\cite{Siegmund:2015:Confounding}} & Study of 385 program comprehension studies with human participants (2001--2010). Focus on confounding factors. & We build on their categorization of confounders in our threats to validity analysis, which is only one of many design aspects we look at.\\
\bottomrule
\end{tabularx}
\end{table}

In addition to the secondary studies described above, several primary studies address specific aspects of the design of code comprehension studies and provide evidence for consequences of certain design decisions.
This started almost 40 years ago with studies on the differences of certain task designs~\cite{Cook:1984:PreliminaryCloze,Hall:1986:Cloze}.
Over the years, this line of research was extended with studies on the comparability of different comprehension measures~\cite{Yeh:2021:IdentifyingConfusion,Shneiderman:1977:Measuring,Ajami:2018:Syntax,Rajlich:1997:TowardsStandards}, the influence of cognitive biases on subjective and objective code comprehension measures~\cite{Wyrich:2021:Mind,Wyrich:2022:Anchoring}, and the actual influence of suspected confounding variables like intelligence, expertise, or code length~\cite{Wagner:2021:Intelligence,Fix:1993:Mental,Teasley:1994:EffectsExpertise,Borstler:2016:RoleMethodChains}.
All of these studies finally culminated in the current discussions about capturing psycho-physiological data to better understand program comprehension~\cite{Fakhoury:2018:ObjectiveMeasures,Siegmund:2020:Studying,Vieira:2021:Psychophysio}.
While these primary studies all contribute to informed decision-making on individual design aspects, our mapping study reviews multiple design characteristics in their entirety and examines interactions between individual aspects.

In summary, designing code comprehension studies has interested the research community for many decades, and yet a detailed, comprehensive study to capture the actual state of past and current study characteristics is lacking.
We fill this gap with a systematic mapping study.

%
% -------------------------------------------------------------------------- %
% Methodology
% -------------------------------------------------------------------------- %
%

\section{Methodology}
\label{sec:methodology}

A systematic literature review (SLR) focuses on synthesizing the available evidence for a topic~\cite{Kitchenham2007}.
Conversely, a systematic mapping study (SMS) is a different type of secondary study with a focus on summarizing the conducted research on a certain topic to provide structure~\cite{Petersen:2008:SystematicMapping}.
Unlike in SLRs, quality-based exclusion is therefore usually not important in an SMS~\cite{Petersen:2015:GuidelinesSystematicMapping}, as the goal is to describe and map what topics were investigated in what way.
Since this is well suited to our objective, we decided to follow the SMS methodology.
The research questions that guided us in our study on experiments about bottom-up source code comprehension are as follows:

\begin{rqs}[leftmargin=*]
    \item How can the studies be classified in terms of their relationship to code comprehension?
    \item What are differences and similarities in the design characteristics of the studies?
    \item What are differences and similarities in the reporting characteristics of the studies?
    \item Which issues and opportunities for improvement are evident in the design and reporting of the studies?
\end{rqs}

The answer to RQ1 is intended to provide insights into what research concerns the primary studies address.
The focus, however, is not on a possible synthesis of evidence, but on the classification of the research topics as well as their relationship to the code comprehension construct, e.g., whether a certain influence on code comprehension was investigated or, conversely, whether the influence of code comprehension on a certain construct was investigated.

\begin{figure*}[htbp]
    \centering
    \includegraphics[width=1\textwidth]{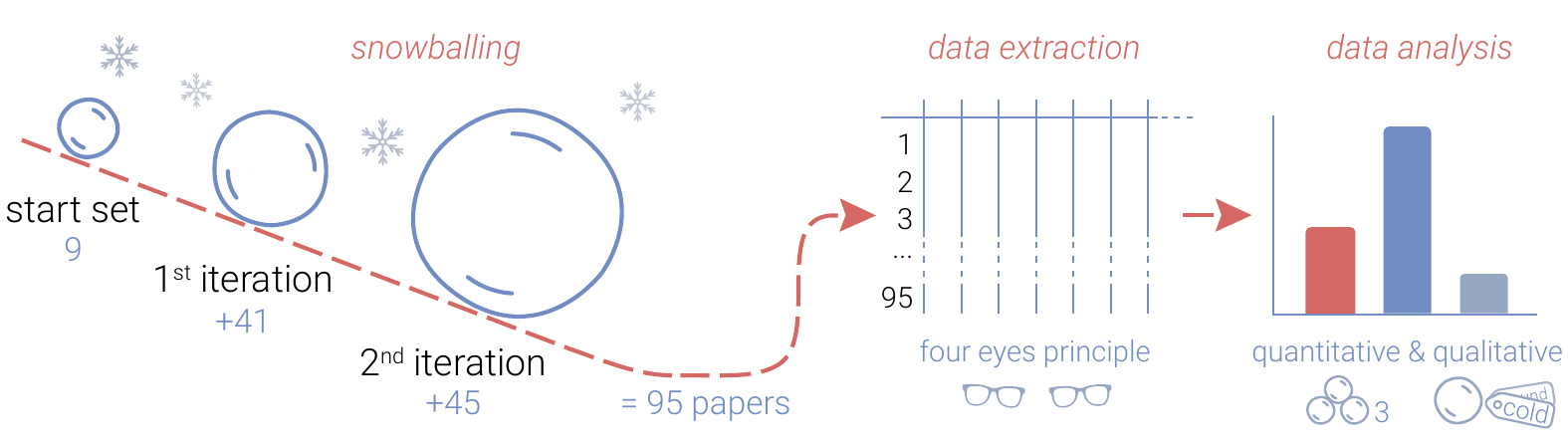}
    \caption{Schematic representation of the research methodology}
    \label{fig:methods}
\end{figure*}

RQ2 forms the core of this work. The answer to this question provides comprehensive insights into the unclear status quo of the design of code comprehension studies.
RQ1--RQ3 were systematically addressed using the methodology described in the following subsections.
The data obtained is analyzed in Section~\ref{sec:results} and forms the basis for a subsequent discussion of RQ4 in Section~\ref{sec:discussion}.
This discussion should improve the quality of code comprehension studies in the medium term and should immediately result in concrete proposals for improving the design and reporting of code comprehension studies.
The following subsections explain our approach to searching and selecting relevant literature, as well as to data extraction and analysis.
An overview of the methodology is depicted in Fig.~\ref{fig:methods}.

\subsection{Search and Selection Process}

We designed our search strategy as a pure snowballing approach, following Wohlin's guidelines~\cite{Wohlin:2014:GuidelinesSnowballing}.
Beginning with a start set of relevant papers known to us, the referenced as well as referencing literature was reviewed for a set of inclusion and exclusion criteria.
This process is called backward and forward snowballing, and it is repeated on newly included literature.
We stopped the literature search after two iterations of backward and forward snowballing, as we approached saturation.

Snowballing is known to researchers who are familiar with systematic literature studies.
Often, however, a database search with predefined search strings is performed before snowballing to identify an initial set of relevant papers.
We refrained from doing so, as a thematically broad literature search via databases leads to a high number of non-relevant search results and thus to considerable effort.
Searching for all papers that measure bottom-up source code comprehension in some way would have to cover a wide range of research topics and would lead to very generic and extensive search terms.
We consider snowballing with a reasonable start set to be more targeted and efficient.
Furthermore, there is evidence that snowballing is similarly effective as and usually more efficient than database searches~\cite{Jalali2012,Wohlin:2014:GuidelinesSnowballing,Badampudi:2015:ExperiencesSnowballing}.

\subsubsection{Start Set}

We build our tentative start set on the results of a recent literature search by~\citet{Munoz:2020:CogCompl}, who searched for program comprehension data sets at the source code level.
Similar to ours, their study searched for empirical studies that measured source code comprehension with human participants, but in a final step filtered for those that had published their dataset.
They then continued working with these data sets to validate a code comprehensibility metric.
\citet{Munoz:2020:CogCompl} identified a total of 10 code comprehension datasets, whose associated papers we use as tentative start set for our study.
Nine of the ten papers meet our inclusion criteria (see Section~\ref{subsec:incluexclu}).
The one excluded paper was a pilot study for another study in the start set and was therefore considered a duplicate.
Wohlin stresses that \enquote{there is no silver bullet for identifying a good start set}~\cite{Wohlin:2014:GuidelinesSnowballing}.
However, he describes characteristics that a good start set should meet, and we will briefly discuss them.

Since~\citet{Munoz:2020:CogCompl} conducted a broad database search to find relevant literature, the start set fulfills the criterion of covering different communities and the criterion of containing keywords that are closely related to our research questions. 
The titles of the papers cover a variety of potential clusters related to program comprehension: program comprehension in general, code and software readability, physiological measures related to program comprehension, code understandability, and code complexity.

A further criterion requires that the start set covers different publishers, publication years, and authors.
The nine papers in the start set were published in IEEE TSE (5), ACM ESEC/FSE (2), and Springer EMSE (2), between 2003 and 2019.
Three of the journal papers are extensions of previously published research papers at the International Conferences on Automated Software Engineering (ASE), Software Analysis, Evolution and Reengineering (SANER), and Program Comprehension (ICPC).
The nine papers were authored by 39 distinct authors.

Finally, a good start set \enquote{should not be too small}~\cite{Wohlin:2014:GuidelinesSnowballing}.
While nine papers seems to be a small number, the papers in the start set have been cited about 400 times and refer to about 500 papers.
Hence, already after the first snowballing iteration, we will have screened about 900 papers.

\subsubsection{Inclusion and Exclusion Criteria}
\label{subsec:incluexclu}
During the search process, a paper was included in the final dataset only if it met all the inclusion criteria (I) and none of the exclusion criteria (E):

\begin{itemize}
    \item[\textbf{I1~}] Reports an empirical study with human participants.
    \item[\textbf{I2~}] Measures bottom-up source code comprehension.
    \item[\textbf{I3~}] Published before 2020.
    \item[\textbf{I4~}] Published in a peer-reviewed journal, conference, or workshop.
    \item[\textbf{E1~}] Is not available in English.
    \item[\textbf{E2~}] Is a meta-study on code comprehension experiments.
    \item[\textbf{E3~}] Is a replication without substantial modification.
    \item[\textbf{E4~}] Is a duplicate or extension of an already included paper.
\end{itemize}

\noindent
Inclusion criteria I1 and I2 ensured that a paper was within the scope of this work.
Examples of code comprehension studies that do not meet I1 include those that only analyze existing data (see, e.g.,~\cite{Posnett:2011:Simpler,Trockman:2018:automaticallyreanalyzed}).
I2 was only fulfilled if the quality or performance of a participant in understanding code was measured (not, for example, \textit{how} participants proceed in understanding code, as in, e.g.,~\cite{Latoza:2007:program,Arunachalam:1996:cognitive}).
Of all the criteria, I2 represents the one that was the most challenging to evaluate because primary studies to date rarely define the construct under investigation, and we will later see that authors use different terms for the same construct.
To evaluate this inclusion criterion, we read the entire paper unless it could already be clearly excluded based on the abstract.
We also emphasize that in evaluating I2, we focused on the authors' intent and only rarely judged at this stage of our study whether the chosen task design of a primary study can actually measure understanding.
If we could not find any explicit indication that the primary study was intended to measure code comprehension and at the same time the task design was not explicit in this respect, we did not include the primary study. 

The rationale for restricting our search to papers published before 2020 (I3) was that we started our literature search in early 2020 and that this constraint would increase the reproducibility and extensibility of our approach.
Including only peer-reviewed literature (I4) served the purpose of quality assurance.
Thus, our dataset contains only primary studies whose methodology has been considered sound by at least a few members of the community. 

A paper was excluded if it was inaccessible to us due to a language barrier (E1), if it was a meta-study (E2), or if the design of the study was similar to that of an already included one because it was a replication, duplicate, or extension (E3 and E4).
We added E3 and E4 to avoid possible bias in the quantitative analysis of the data.
Even with extensions, the original study designs are rarely modified, which means that inclusion of the original study and the extension would cause the specific design decisions to be considered twice in the analysis.
Regarding the impact of E3, we identified only one replication that would have met our inclusion criteria, namely~\citet{Fucci:2019:Replication}.

The set of papers resulting from the search and selection process is characterized in detail in Section~\ref{subsec:included}.
In total, we included 95 papers. The 9 papers in the start set led us to 41 relevant papers.
On these 41 papers, we again performed a snowballing iteration, which yielded 45 more relevant papers.
The ratio of newly included literature to literature screened in this second snowballing iteration was 1.2\% (compared to 4.6\% in the first iteration), which made us confident to assume a saturated data set at this point. We therefore stopped searching for relevant literature after the second iteration.

\subsection{Data Extraction and Analysis}

Before data extraction, all data items to be extracted were defined with a description as precise as possible.
In addition to DOI, APA citation, publication year, and venue, 17 additional columns were filled in a spreadsheet for each of the 95 included papers to characterize different aspects of the study design.
The extracted data items are listed in Table~\ref{tab:data_items}.
The results are presented according to the same data item categories in Section~\ref{sec:results} under subsections of identical names.
The holistic results for RQ1--RQ3 are then used to answer RQ4 in the Discussion section.

\begin{table}[th]
\caption{Extracted data items}
\label{tab:data_items}
\begin{tabularx}{\columnwidth}{l X l}
\toprule
Category & Data Items & Used to Answer \\
\midrule
Included Papers & DOI, APA citation, publication year, venue &\\
Study Themes & study category & RQ1\\
Construct Naming & construct name & RQ3\\
Study Designs & research method and experiment design & RQ2, RQ3\\
\makecell{Participant\\~~Demographics} & \#participants, demographics & RQ2, RQ3\\
Setting and Materials & \#snippets per participant, snippet pool size, code snippet source, snippet selection criteria, remote vs. onsite, paper vs. screen, programming languages & RQ2, RQ3\\
\makecell{Comprehension Tasks\\~~and Measures} & comprehension tasks, construct measures & RQ2, RQ3\\
Reported Limitations & limitations & RQ2, RQ3\\
\bottomrule
\end{tabularx}
\end{table}

The first two authors independently performed the extraction for all papers in the start set (9) and the first snowballing iteration (41).
In several meetings, each extracted cell was then reviewed for agreement and, if necessary, a consensus was reached through discussion.
Over time, we thus established a strong understanding of the literature and difficulties in the extraction process.
This four-eyes principle was highly time-consuming, but led to greater certainty that the papers examined were correctly understood and that no relevant details were overlooked.

In the second snowballing iteration, each of the first two authors extracted half of the 45 newly added papers.
Items about which one was unsure were subsequently examined by the other.
The same applied to papers where one of us came to the conclusion during extraction that the study should actually be excluded.
Such a decision was never made alone.
In the supplemental materials, we transparently indicate which papers were extracted by which author(s).

After extracting the design characteristics for all 95 papers, the data were analyzed both quantitatively and qualitatively.
Some data items such as publication date, number of study participants, and used programming language could for the most part be analyzed without further data transformation.
Other data items such as the study themes and comprehension tasks first had to be classified using thematic analysis~\cite{Cruzes2011}, since the diversity of these data was unknown at the time of data extraction.
Finally, there were items for which we extracted descriptive quotations that we first labeled and then analyzed by category building.
Examples of such data items are the criteria for code snippet selection and threats to validity discussed by the authors of the primary studies.

\subsection{Data Availability}
\label{subsec:dataset}
Following open science principles~\cite{Mendez:2020:Openscience}, we publicly disclose the dataset of 95 studies together with the extracted data items~\cite{zenodo:dataset}.
The dataset\footnote{\url{https://zenodo.org/record/6657640}} details which studies were included in which step and which studies were extracted by which authors.
In addition, we publish a large part of our data analyses, which can be used, e.g., to trace how characteristics of each individual study design were labeled and categorized in specific analyses.

\subsection{Threats to Validity}
Our methodological approach has a few limitations that are relevant for the interpretation of the results and that we would therefore like to point out in advance.

In the \textbf{search and selection process}, we restricted ourselves to papers published before 2020.
This was to maintain replicability, as our systematic literature search began in 2020.
Our results will show a trend that a large proportion of included papers were published in recent years.
Accordingly, we should assume that some relevant papers were also published in 2021 and 2022, which may have different design characteristics. 
We consider it worthwhile to conduct a similar mapping study again as early as 2026 to capture the latest trends in designing code comprehension studies.

Inclusion criterion I2 required a subjective judgment on whether a primary study intended to measure bottom-up source code comprehension. 
We prioritized avoiding false positive inclusion decisions over avoiding false negative ones.
We aimed for a dataset that included only studies meeting our criteria with a high level of certainty, possibly resulting in the incorrect exclusion of some papers.
On a related note, no search process can claim complete coverage of all relevant studies, whether it is a database search, snowballing, or a hybrid approach.
Nevertheless, we are confident that the 95 papers found are a representative set of papers for our inclusion criteria.

Regarding the quality of individual primary studies, we relied solely on inclusion criterion I4, i.e., that the study was peer-reviewed.
We did not perform any further quality assessment, as sometimes suggested in guidelines for systematic literature reviews~\cite{Keele:2007:Guidelines}.
This decision is based on our motivation to obtain the status quo of study designs accepted by the community.
It is therefore unavoidable (and even \enquote{desirable} for us) that some primary studies included may have design or reporting flaws.
After all, it is these weaknesses that we aim to systematically identify and discuss.

As for the process of \textbf{data extraction and analysis}, we would like to point out that in the second snowballing iteration, data extractions were mostly performed by individual researchers.
The same applies to the analysis of extracted data.
The categories of extracted data items were generally analyzed and proposed by a single researcher.
In both cases, however, results were discussed in detail, and the opinions of the other authors were regularly solicited.
Each author also worked with data points extracted from the other authors in the data analysis, which in a sense verified extracted data points, since one had to occasionally look at the papers again for their context.

Additionally, a few papers describe several different experiments.
In case of differences between these experiment facets, we therefore either selected the dominant variant (e.g., for the experiment factor design) or a suitable aggregated value (e.g., median number of snippets per participant) during the synthesis.
As described in Section~\ref{subsec:dataset}, we make the entire process as transparent and traceable as possible by publishing the data and analyses.

%
% -------------------------------------------------------------------------- %
% Results
% -------------------------------------------------------------------------- %
%

\section{Results}
\label{sec:results}

We present the results grouped by the categories introduced in Table~\ref{tab:data_items}.
A critical discussion of findings of particular interest follows separately in the subsequent discussion section~\ref{sec:discussion}.

\subsection{Included Papers}
\label{subsec:included}

\begin{figure*}[h]
    \center
    \includegraphics[width=1\textwidth]{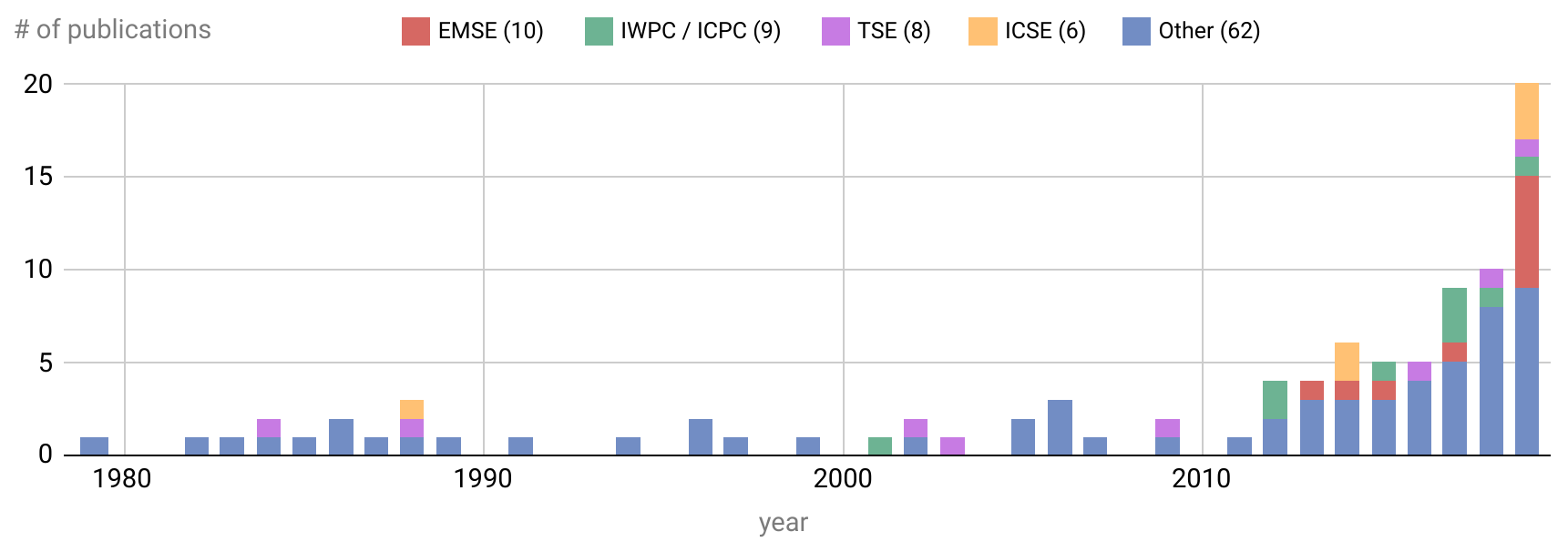}
    \caption{Number of publications per year and venue}
    \label{fig:includedPapers}
\end{figure*}

We included a total of 95 primary studies published from 1979 to 2019.
Fig.~\ref{fig:includedPapers} shows the number of published studies per year.
A large proportion of the included studies were published between 2010 and 2019 (67.4\%), and about half of all included studies were published from 2015 and onward (51.6\%).
Forty-one papers were published in a journal, 45 in conference proceedings, and 9 in workshop proceedings.
The papers were published in 51 different venues, of which 39 appeared only once in the dataset.
The top 4 venues by number of included papers are EMSE (10), IWPC/ICPC\footnote{The International Conference on Program Comprehension (ICPC) was a workshop (IWPC) until 2005. One included paper was published at IWPC.} (9), TSE (8), and ICSE (6).
From the 62 \enquote{Other} papers depicted in Fig.~\ref{fig:includedPapers}, 39 appear in venues from which we only included one paper each.
The remaining 23 \enquote{Other} papers appear in the following eight venues:  ESEC/FSE (4), International Journal of Man-Machine Studies (4), EMIP (3), ETRA (3), International Journal of Human-Computer Studies (3), CHI (2), ICSME (2), PPIG (2).

\subsection{Study Themes}
\label{subsec:studythemes}
As a general starting point for study themes, we were interested in the study distribution according to the following three relationship types:

\begin{itemize}
    \item \textbf{X $\Rightarrow$ code comprehension:} the study analyzes how one or more other constructs (X) influence code comprehension, e.g., identifier naming length or developer experience.
    \item \textbf{X $\Leftrightarrow$ code comprehension:} the study analyzes the correlation between code comprehension and one or more other constructs (X), e.g., source code metrics or eye tracking data.
    \item \textbf{code comprehension $\Rightarrow$ X:} the study analyzes how code comprehension influences one or more other constructs (X), e.g., developer motivation or happiness.
\end{itemize}

A study could be mapped to several of such instances, depending on its scope and research questions.
For example, \citeauthor{S1} studied three instances in S1:
\textit{fMRI data $\Leftrightarrow$ code comprehension} (RQ1), \textit{code complexity $\Rightarrow$ code comprehension} (RQ2), and \textit{programming experience $\Rightarrow$ code comprehension} (RQ3).
Nonetheless, 78 of 95 studies in our sample were mapped to a single relationship instance (82\%), while only 13 and 4 studies were mapped to 2 and 3 instances respectively.
No study in our sample produced more than three instances.
To analyze the distribution of relationship types, we assigned a primary type to the 17 studies with more than one type.
Each of these 17 papers was carefully analyzed to identify if one of the RQs might be the central focus of the paper, with the remaining RQs being either preliminary or smaller side investigations.
In such cases, the relationship type of the central RQ was used.
For example, S1 was categorized as \textit{X $\Leftrightarrow$ code comprehension} because RQ1 was the primary focus of their study (the correlation between code comprehension and fMRI data).
In the rare cases where no primary RQ could be determined, we instead opted for the most frequent relationship type of all RQs of the paper.

As a result, the majority of studies, namely 67 of 95 (71\%), belong to the \textit{X $\Rightarrow$ code comprehension} relationship type, i.e., analyzing which constructs influence comprehension in which way is the most popular type of research in our sample.
The remaining 28 papers all belong to the relationship type \textit{X $\Leftrightarrow$ code comprehension}, i.e., they analyzed correlations between constructs and code comprehension, mainly to evaluate suitable measurement proxies.
Surprisingly, we did not find a single study for the type \textit{code comprehension $\Rightarrow$ X}.
Potential explanations could be that such research is perceived as either not industry-relevant or not suitable until further progress is made in defining and measuring code comprehension as a construct.

\begin{small}
\begin{table*}[ht]
    \caption{Study purpose categories}
    \label{tab:studyPurposeCategories}
    \centering
    \begin{tabularx}{\textwidth}{lrrX}
        \toprule
        Category & \# of Mentions & \# of Labels & Description (\enquote{One study goal is to analyze\ldots})  \\
        \midrule
        semantic cues & 30 & 3 & how semantic information in source files (e.g., identifiers or comments) influence comprehension.  \\
        \makecell{developer\\ characteristics} & 30 & 5 & characteristics of developers in the context of comprehension (e.g., experience or code familiarity).  \\
        \makecell{psycho-physiological\\ measures} & 29 & 7 & the feasibility of psycho-physiological measures collected from developers (e.g., fMRI or EEG) as proxies for comprehension.  \\
        code structure & 28 & 10 & how structural attributes of source files (e.g., control structures or code size) influence comprehension.  \\
        code evaluation & 10 & 2 & ways to evaluate code understandability (e.g., metrics or self-assessment).  \\
        \makecell{programming\\ paradigms} & 9 & 4 & programming paradigms w.r.t. comprehension (e.g., object orientation or reactive programming).  \\
        \makecell{comprehension\\ support} & 9 & 4 & ways to better understand code without changing it (e.g., code browsing techniques or diagrams).  \\
        visual characteristics & 8 & 3 & the influence of visual code characteristics on comprehension (e.g., indentation or syntax highlighting).  \\
        code improvement & 4 & 3 & ways to change existing code towards better understandability (e.g., methods or tools).  \\
        test code & 3 & 1 & ways to improve test code w.r.t. comprehension.  \\
        mental models & 3 & 1 & how developers mentally represent code and how this influences comprehension. \\
        \bottomrule
    \end{tabularx}
\end{table*}
\end{small}

\begin{figure}[H]
    \center
    \includegraphics[width=0.67\columnwidth]{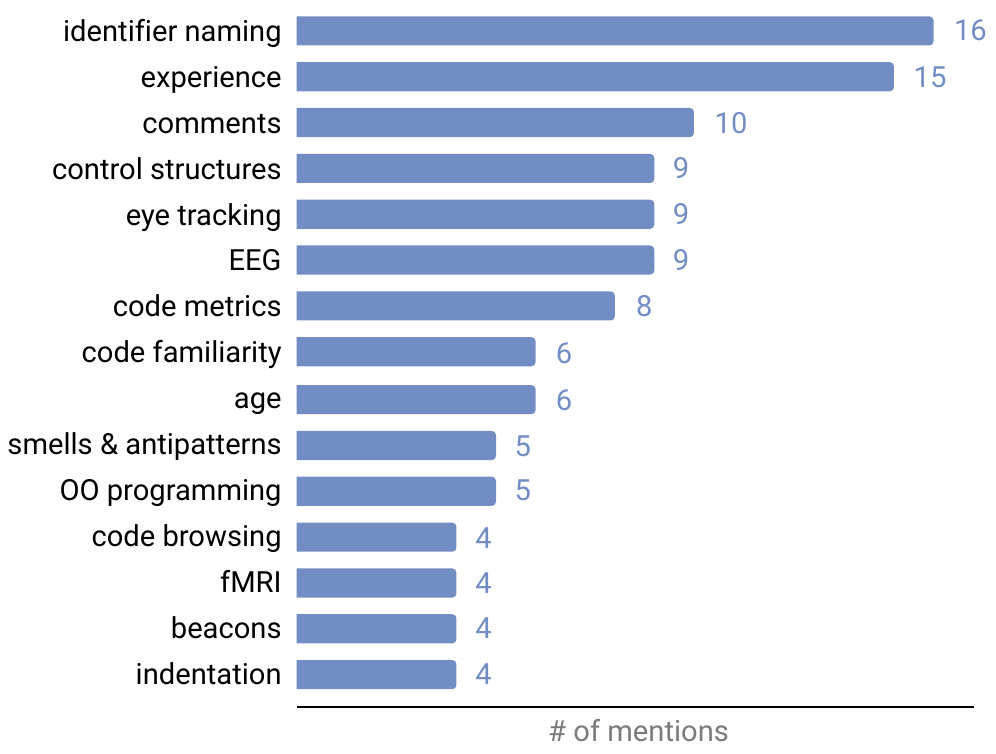}
    \caption{Most popular study themes (assigned to at least 4 papers)}
    \label{fig:studyThemes}
\end{figure}

In addition to these general relationship types, we also assigned one or more thematic labels to each paper based on the study purpose (between 1 and 10 labels, median of 1).
Afterwards, these labels were grouped into higher-level categories (see Table~\ref{tab:studyPurposeCategories}).
In total, we created 43 labels assigned to 11 categories.
Categories consisted of between 1 and 10 labels (median of 3).
The most popular categories in our sample were \textit{semantic cues} (e.g., identifier naming or comments), \textit{developer characteristics} (e.g., experience or age), \textit{psycho-physiological measures} (e.g., eye tracking or EEG), and \textit{code structure} (e.g., control structures or procedure usage).
Less studied areas were \textit{code improvement}, \textit{test code}, and \textit{mental models}.
Concerning individual labels, 15 of the 43 themes were assigned to at least 4 papers (see Fig.~\ref{fig:studyThemes}).
Popular themes included \textit{identifier naming} (16 studies), \textit{experience} (15), \textit{comments} (10), and \textit{control structures} (9).
Additionally, three labels from the \textit{psycho-physiological measures} category are among the top 15: \textit{eye tracking} (9), \textit{EEG} (9), and \textit{fMRI} (4).
\newpage

\begin{figure}[tbh]
    \center
    \includegraphics[width=\columnwidth]{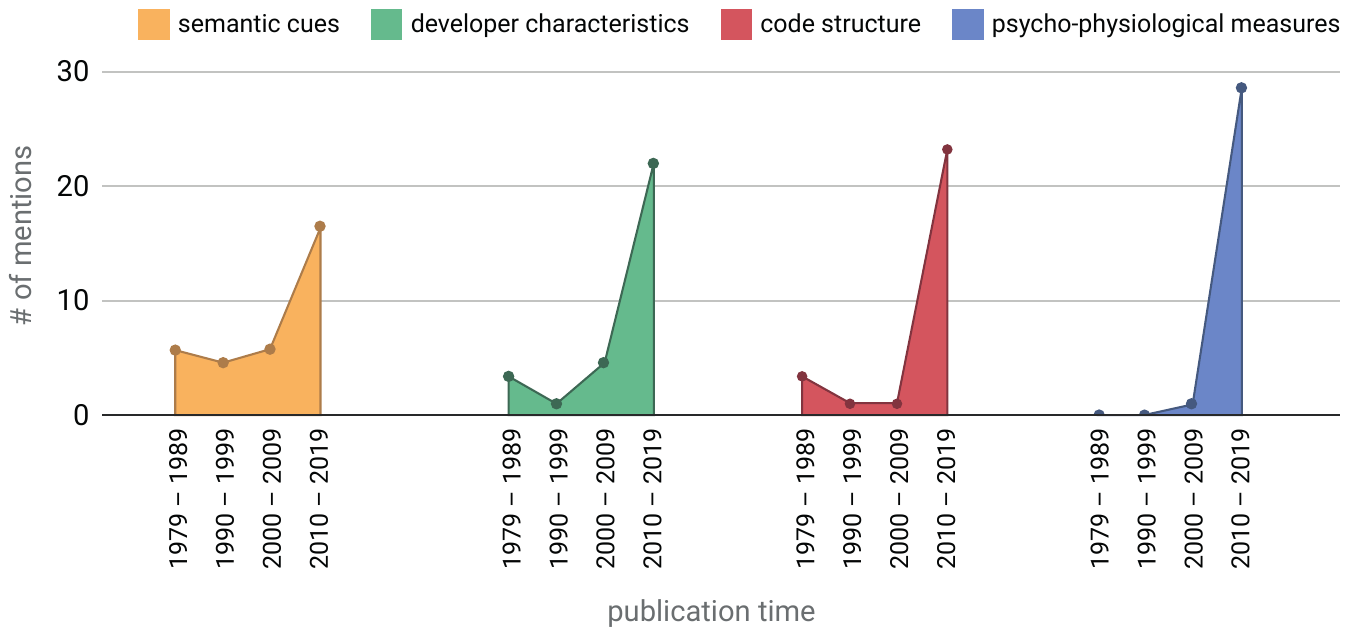}
    \caption{Evolution of the four most popular study categories according to their number of occurrences}
    \label{fig:studyThemeEvolution}
\end{figure}

When analyzing the evolution of the four largest thematic categories over the years (see Fig.~\ref{fig:studyThemeEvolution}), we see that studies on \textit{semantic cues} like identifier naming and code comments were most popular early on but lost their relative market share in the last decade.
The relative popularity of studies on \textit{code structure} and \textit{developer characteristics} increased, even though both are not comparable to the meteoric rise of \textit{psycho-physiological measures}.
Enabled by technological advances and the increased affordability of relevant devices, the last 10 years saw an abundance of studies evaluating data from, e.g., eye tracking, fMRI, EEG, NIRS, or HRV as proxies for code comprehension.
This signifies a shifting focus of the research community towards a more developer-centric and individual perspective of code comprehension.
As an example, Couceiro et al. even envision \enquote{biofeedback code highlighting techniques} in S51, i.e., reporting complex or potentially buggy lines of code in real-time during development, using data from non-intrusive techniques such as HRV, pupillography, and eye tracking.

\subsection*{Answer to RQ1: How can code comprehension experiments be classified in terms of their relationship to code comprehension?}

While the following result subsections provide a combined answer to research questions 2 and 3, we can already provide a synopsis of our first research question at this point.
The majority of studies in our sample investigated how a condition affects code comprehension (71\%).
The remaining papers examined correlations between a condition and code comprehension.
None of the 95 studies, however, looked at the consequences of code comprehension on any other construct.

Thematically, the studies most frequently investigate the influence of semantic cues, developer characteristics, and code structure on code comprehension.
Studies evaluating the feasibility of psycho-physiological measures collected from developers as proxies for comprehension are particularly popular today.

\subsection{Construct Naming}

\begin{figure}[tb]
    \center
    \includegraphics[width=0.7\columnwidth]{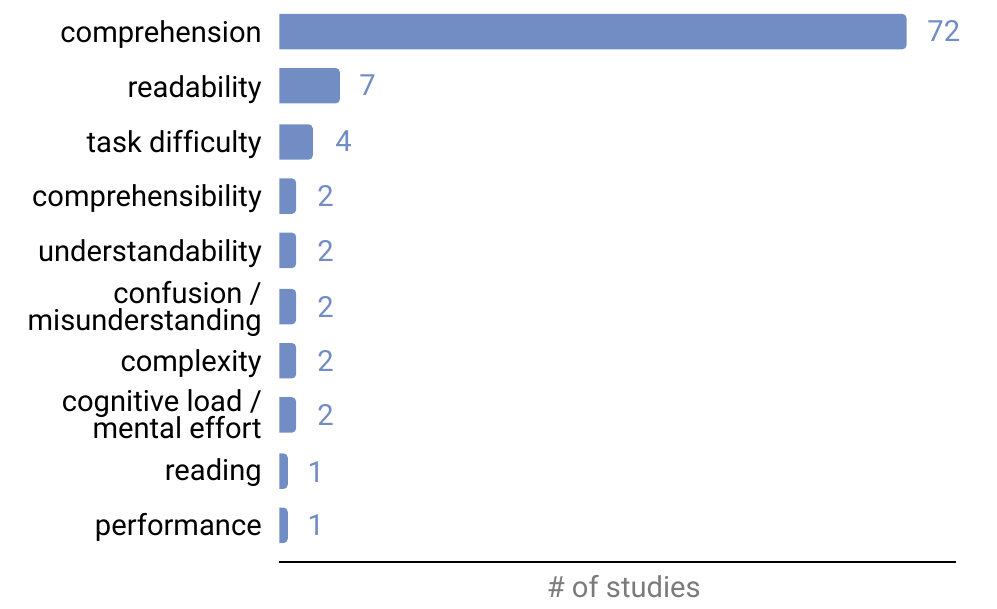}
    \caption{Names for the central construct in our primary studies}
    \label{fig:constructNames}
\end{figure}

For each primary study, we extracted the name used for the central construct under investigation.
In total, we identified 10 different construct names (see Fig.~\ref{fig:constructNames}), which optionally could be combined with \enquote{code} or \enquote{program}.
The most frequently used name was by far \enquote{comprehension}, with 72 of 95 studies (76\%).
Seven studies used \enquote{readability} as the construct name, even though they measured comprehension, i.e., participants' level of understanding and not their ability for visual perception (see Section~\ref{sec:comprehension}).
Four studies focused on their concrete experiment tasks and explicitly used \enquote{task difficulty} as a construct.
The remaining seven constructs were all used by at most two studies.

The majority of studies (73 of 95) also used constructs describing an \textit{activity} like \enquote{comprehension} or \enquote{reading}.
Only 22 studies used an \textit{attribute} like \enquote{comprehensibility}, \enquote{complexity}, or \enquote{cognitive load} as the construct.
Lastly, nearly half of the studies (47 of 95) combined their construct name with \enquote{program}, 24 studies combined it with \enquote{code}, with the remaining 24 studies refraining from using such a modifier.
As an example, 45 of the 72 studies using \enquote{comprehension} as the construct phrase it as \enquote{program comprehension}, 18 as \enquote{code comprehension}, and 9 only as \enquote{comprehension}.
With \enquote{program comprehension} being a superset of \enquote{code comprehension}, using the latter would be more precise, as many studies we excluded for not being concerned with bottom-up source code comprehension used the older term \enquote{program comprehension}.

\subsection{Study Designs}
Concerning the used research methods, 94 of 95 studies conducted some form of \textit{experiment}.
The one exception was S42, a field study performed by \citeauthor{S42} to test a proposed code improvement method.
Since the used terminology on this may differ, we did not distinguish between different types of experimentation, such as \textit{controlled experiments}, \textit{experiments}, or \textit{quasi-experiments}~\cite{Wohlin2012}.
Moreover, such a distinction would have provided little value for answering our RQs.
We therefore classified all studies that manipulated at least one independent variable while controlling for other factors as an \textit{experiment}.
Nonetheless, the degree of manipulation and control differs between studies.
For example, \citeauthor{S8} used two explicit treatments in S8, namely the object-oriented \texttt{Observer} design pattern vs. reactive programming constructs.
In comparison, \citeauthor{S6} did not have such clear treatments in S6.
Instead, their manipulation manifested in the varying complexity of the many code snippets shown to participants.

To be able to analyze trends and differences in experiment designs, we extracted the following properties that are loosely based on the methodological publications from \citet{Juristo2001} and \citet{Wohlin2012}:

\begin{itemize}
    \item \textbf{Factor design:} we extracted whether the experiment used a \textit{factorial} design, i.e., several factors were manipulated in parallel, or a \textit{one-at-a-time} design, i.e., only a single factor was manipulated in parallel.
    \item \textbf{Allocation design:} we extracted whether the experiment used a \textit{within-subject} design (also called \textit{repeated measures} design), i.e., each participant received each treatment at least once, or a \textit{between-subject} design (also called \textit{independent measures} design), i.e., each participant only received a single treatment.
    \item \textbf{Randomization or counterbalancing of tasks}: this is important for \textit{within-subject} designs to combat familiarization and carryover effects, especially for the \textit{within-subject} variant called \textit{crossover} design~\cite{Vegas2016}. We extracted whether this was applied (\textit{yes} or \textit{no}).
    \item \textbf{Experience- or skill-based balancing}: this is especially important for \textit{between-subject} designs, as they do not account for individual differences between participants, e.g., regarding expertise or motivation~\cite{Vegas2016}. Consciously balancing your groups regarding potential confounders can mitigate such effects. We extracted whether this was applied (\textit{yes} or \textit{no}).
\end{itemize}

\begin{figure}[th]
    \center
    \includegraphics[width=0.7\columnwidth]{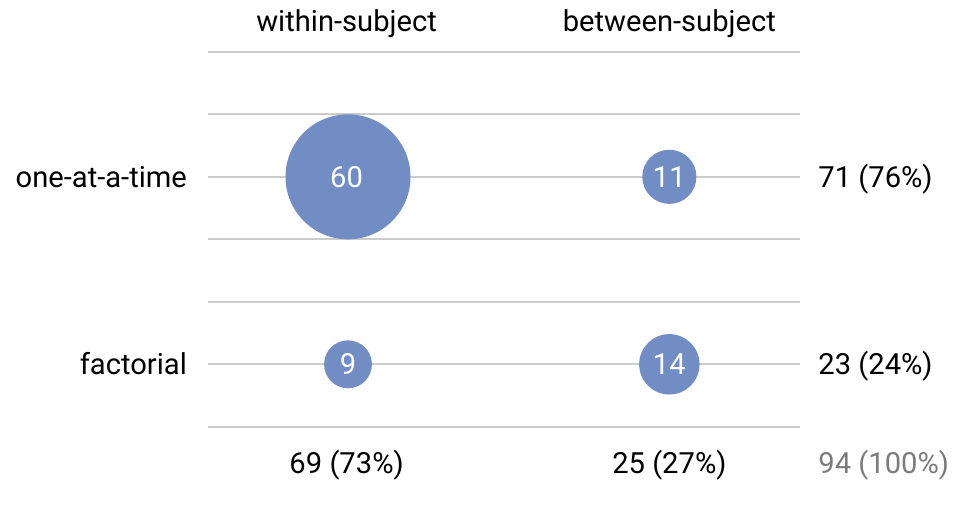}
    \caption{Distribution of factor and allocation design}
    \label{fig:experimentDesign}
\end{figure}

Concerning the factor design, 71 of the 94 experiments in our sample chose a \textit{one-at-a-time} design.
Only 23 studies used a \textit{factorial} design, which is generally considered to be both more complex and powerful.
Concerning allocation design, we see a similar imbalance, with 69 studies choosing a \textit{within-subject} design, which is considered to be more robust against participant heterogeneity.
Only 25 studies relied on a \textit{between-subject} design, where skill differences between groups can be a threat to the validity of the results.
When combining factor and allocation design (see Fig.~\ref{fig:experimentDesign}), \textit{within-subject} is substantially more popular for \textit{one-at-a-time} designs (85\% of studies), while \textit{between-subject} is a bit more frequently chosen for \textit{factorial} designs (61\%).

\begin{figure}[t]
    \center
    \includegraphics[width=\columnwidth]{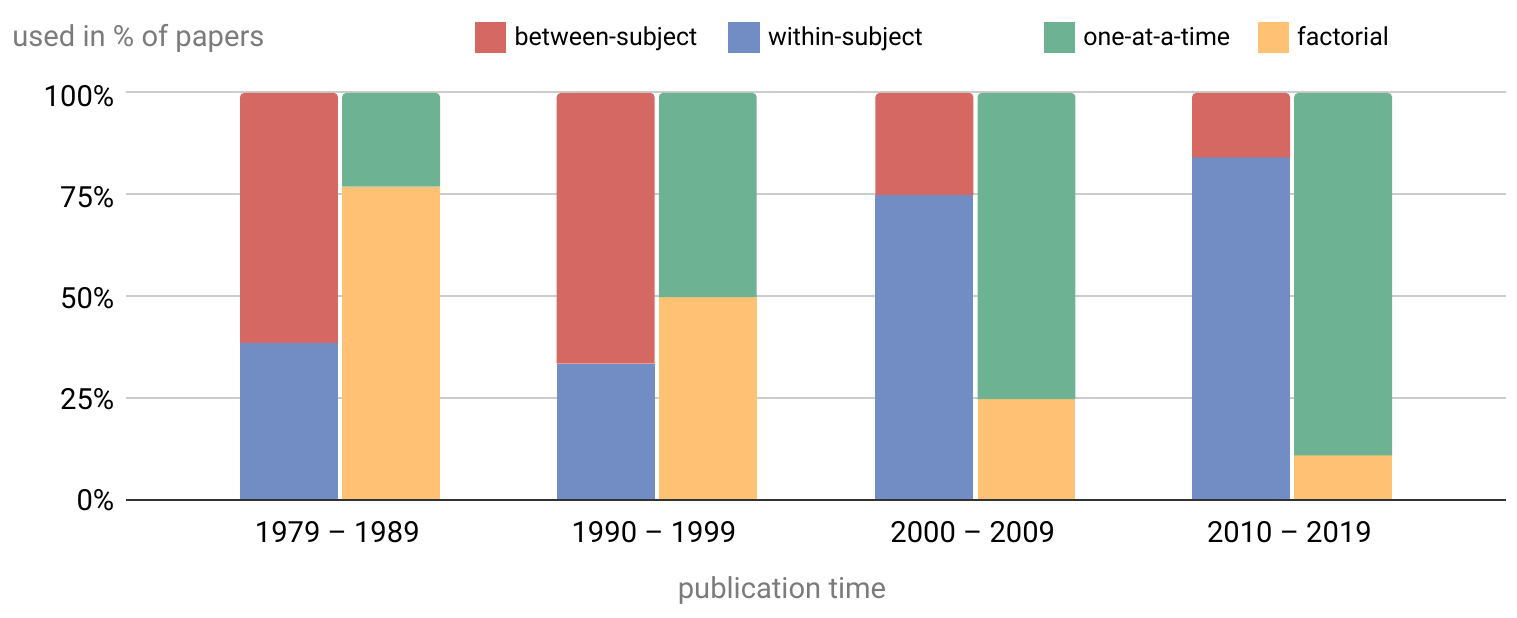}
    \caption{Evolution of factor and allocation design}
    \label{fig:experimentDesignEvo}
\end{figure}

A combined analysis with the publication year also reveals a shift in experiment design over time (see Fig.~\ref{fig:experimentDesignEvo}).
Before 1990, most studies used \textit{factorial between-subject} designs.
A typical example is S67, a 2x2 factorial experiment published by \citeauthor{S67} in 1985.
He analyzed the effect of internal procedures and code comments on comprehension by assigning each participant to one of four groups: no procedures and no comments, no procedures and comments, procedures and no comments, and procedures and comments.
Over the years, \textit{one-at-a-time within-subject} designs started to rise in popularity, until they became the dominant form of code comprehension experiments from 2010 and onwards.
A typical example for this is S3, an experiment to analyze the impact of identifier naming styles on code comprehension published by \citeauthor{S3} in 2019.
In their study, every participant received tasks for each of the three levels of naming styles: single letters, abbreviations, and words.

Lastly, we analyzed the frequency of applying suitable precautions for the respective weakness of the two types of allocation design, i.e., if \textit{within-subject} designs used randomization/counterbalancing and if \textit{between-subject} designs used experience- or skill-based balancing during group assignment.
Of the 69 \textit{within-subject} designs in our sample, 45 applied randomization or counterbalancing of tasks (65\%).
This means that approximately one third did not use this proven technique to counteract potential familiarization effects.
However, the numbers are much worse for \textit{between-subject} designs.
Of the 25 experiments with this allocation design, only 6 balanced their groups based on experience or skill (24\%), i.e., 76\% took insufficient or no precautions in this area.

\subsubsection*{Summary}
Most experiments in our sample preferred a one-at-a-time design over a factorial design. In addition, most studies have opted for a within-subject design rather than a between-subject design. 
However, we noted that preferences regarding factor and allocation design have changed over time: between-subject and one-at-a-time experiments characterized the majority of experiments 30 to 40 years ago.

\subsection{Participant Demographics}
\begin{figure*}[h]
    \center
    \includegraphics[width=1\textwidth]{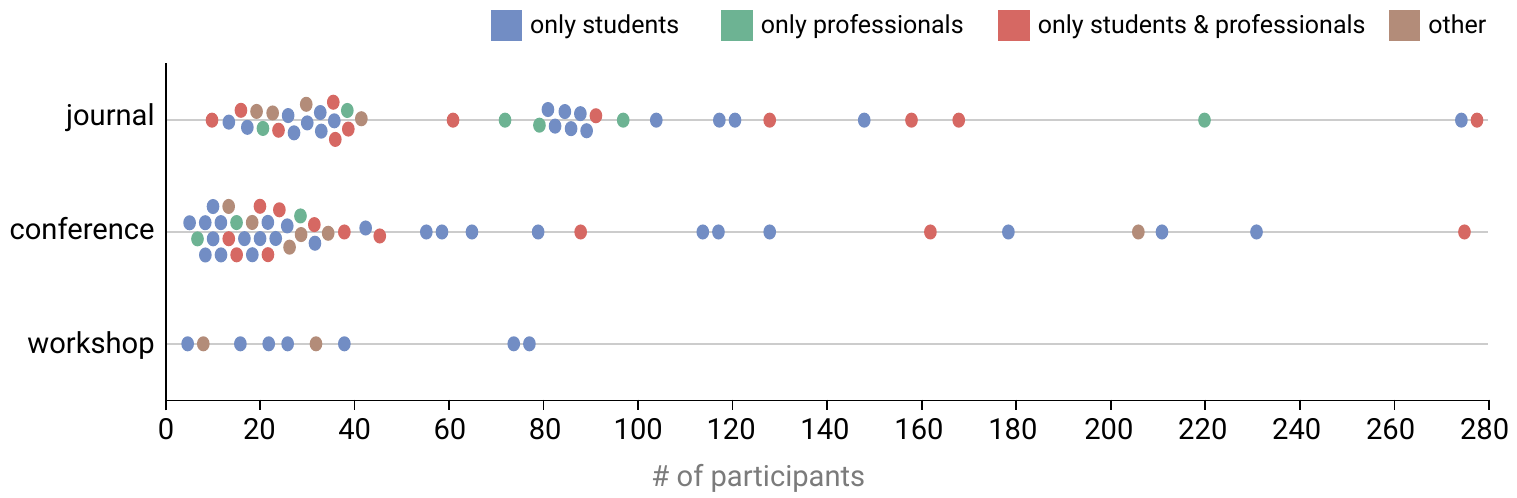}
    \caption{Number of participants per study, grouped by venue type. Colors indicate the demographic composition of the sample}
    \label{fig:participants}
\end{figure*}

All included studies have in common that they used a sample of human participants to answer their research question.
The number of participants ranges from 5 to 277, with a median of 34.
Fig.~\ref{fig:participants} shows the distribution of sample size per paper.
For papers that reported different numbers of participants, for example, due to multiple reported sub-experiments, the mean of minimum and maximum reported number of participants was used for that paper (this was the case for 13 papers).

About half of the 95 papers (53.7\%) reported a sample that consisted entirely of students.
For nine papers (9.5\%), only professionals were included in the sample.
In 24.2\% of the samples, the participants were composed of students and professionals.
Seven papers (7.4\%) recruited a sample of students and faculty members.
For four papers (4.2\%), we were either unable to find any information at all or only very vague information on participant characteristics.

Looking at the number of participants and their demographic characteristics by type of venue (journal/conference/workshop), substantial differences emerge.
The number of participants for journal articles with a median of 61 (mean: 77.5) is higher than the median of 26 for conferences (mean: 58.2) and the median of 26 for workshops (mean: 33.1). Furthermore, a sample in a journal paper is least likely to be composed entirely of students (46.3\% of papers compared to 55.6\% for conferences and 77.8\% for workshops).
None of the nine included workshop papers sampled any professionals.
Two of the workshop papers recruited faculty members in addition to students.

\subsection{Setting and Materials}
We also extracted and analyzed several aspects of the experiment settings and used materials, i.e., the code snippets.

\subsubsection{Experimental Settings}

Regarding the \textit{experiment location}, 72 of the 95 studies were conducted onsite (76\%) and therefore had strong control of the study environment.
In 14 cases, a remote experiment was chosen, thereby trading off some control for more and a wider variety of participants.
Two studies adopted a mixed approach, with both onsite and remote participation.
For seven studies, we could not reliably determine the study location from the paper.
The earliest remote study (S24 by \citeauthor{S24}) was already conducted in 2007 based on a browser-based Java applet, but the median year was 2019.
Due to the COVID-19 pandemic, we expect the percentage of remote experiments published in 2020 and later to be substantially larger.

As the \textit{snippet medium}, 67 studies presented the code on a screen (71\%), while 21 studies used code printed on paper.
One study used both mediums.
In six cases, we could not reliably determine if a screen or paper was used.
While we found studies published as recent as 2018 using paper (S34 by \citeauthor{S34}), the median year was 1996, indicating that this becomes increasingly rare.
Unsurprisingly, location and medium are also interrelated to some degree, i.e., all 14 remote experiments used a screen, while all 21 paper-based studies were conducted onsite.

\subsubsection{Experimental Materials}

\begin{figure}[th]
    \center
    \includegraphics[width=0.7\columnwidth]{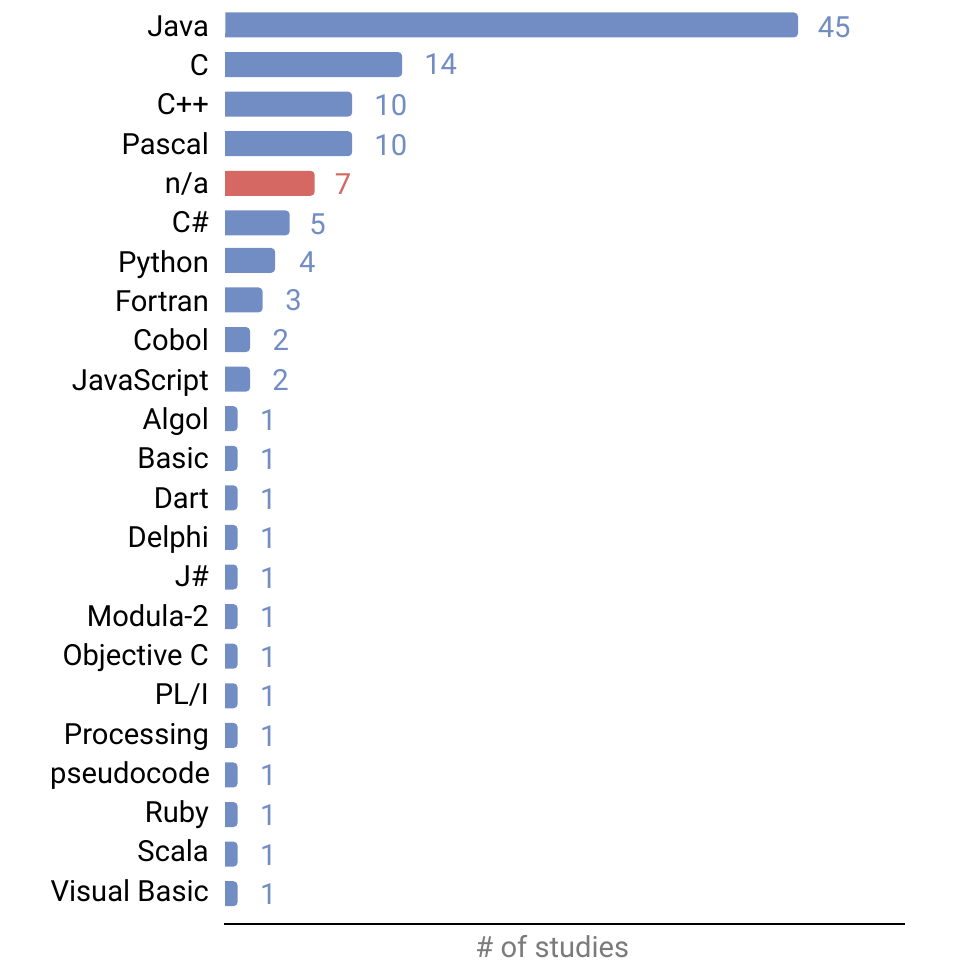}
    \caption{Programming languages used in the experiments}
    \label{fig:programmingLanguages}
\end{figure}

Concerning the used \textit{programming languages} for the snippets (see Fig.~\ref{fig:programmingLanguages}), we identified a total of 21 unique languages (plus pseudocode).
The majority of studies (84 of 95) only used a single language for their snippets, with eight studies using two languages, and a single study using three, four, and nine languages respectively.
The latter outlier is S42, where \citeauthor{S42} let participants bring their own code snippets, resulting in a variety of languages.
With 45 papers (47\%), Java is by far the most used language in our sample.
It is followed by C (14 studies), C++ (10), and Pascal (10).
For seven papers, we could not reliably extract the used programming language from the paper (label \texttt{n/a}).
A total of 13 languages were only used in a single study.
Studies using older languages like Fortran, Cobol, Pascal, Algol, Basic, PL/I, or Modula-2 were all published before 2000, with a median year of 1987.

We also extracted and analyzed the \textit{snippet pool size} and \textit{number of snippets per participant} for each study.
In four cases, we could not reliably determine these metrics from the paper.
The remaining 91 studies had a median snippet pool size of 9, with a mean of $24.7$ over the range from 1 to 389 snippets.
We found that 34 studies (37\%) had 5 or fewer snippets, and 18 studies (20\%) had between 6 and 10 snippets in the snippet pool.
Apart from that, we see a decent variety of snippet pool sizes, with some outliers in the lower hundreds, like S89 by \citeauthor{S89} with 298 and S41 by \citeauthor{S41} with 389 snippets.
In 45 cases, participants worked on the complete snippet pool (100\% snippet percentage).
The majority of these were \textit{within-subject} designs, as treatments were most often embedded into the snippets.
In the remaining studies, participants worked on very different percentages of the snippets, with peaks at 50\%, 33\%, and 25\%.
These correspond to frequently chosen numbers for groups or snippet versions, namely two, three, and four.
With larger snippet pools (30 and more), the snippet percentage tended to be lower, even when no groups or snippet versions were used, most likely to not exhaust participants.
However, there were also several exceptions, e.g., both \citeauthor{S6} (S6, 100 snippets) and \citeauthor{S41} (S41, 389 snippets) let each participant work on all snippets.
In summary, the average study in our sample used 25 snippets and let participants work on 70\% of them (median: 9 snippets with 90\% snippet participation).

\begin{figure}[t]
    \center
    \includegraphics[width=0.7\columnwidth]{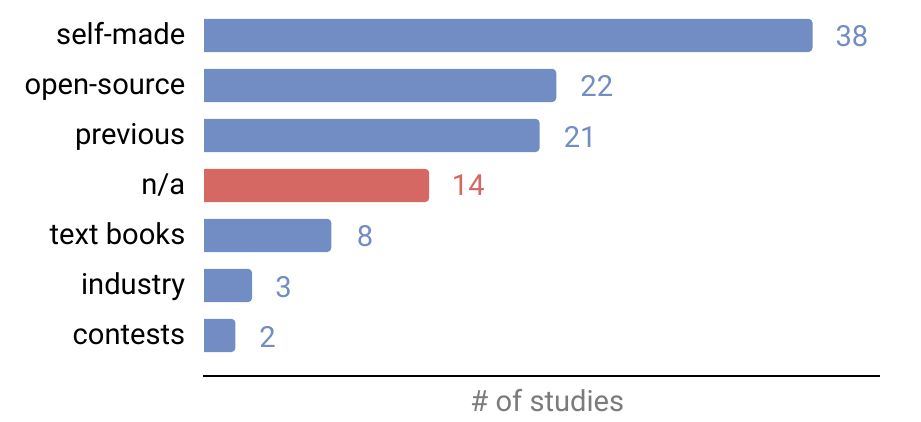}
    \caption{Sources for code snippets}
    \label{fig:snippetSource}
\end{figure}

We further analyzed where the used snippets originated from, i.e., the \textit{snippet source} (see Fig.~\ref{fig:snippetSource}).
For 14 studies, we could not reliably determine the source (\texttt{n/a}).
The majority of the remaining papers used a single source (68 studies), with 13 studies having two sources for their snippets.
The most popular source was that researchers created snippets themselves (38 studies).
In 22 studies, parts of open-source software were used, sometimes slightly adapted, 21 papers simply reused snippets from previous studies, and 8 papers relied on textbooks, e.g., from computer science education.
Only three studies used closed-source code from industry.
Lastly, two papers sourced from programming contests.
Overall, the majority of snippets were self-made or from open-source software, with snippets from proprietary industry projects playing no major role.

\subsubsection{Snippet Selection Criteria}

Finally, we extracted the \textit{snippet criteria}, i.e., the requirements researchers had for the selection or creation of code snippets and how they motivated snippet usage.
In total, we identified 17 criteria (see Table~\ref{tab:snippetCriteria}).
Studies were labeled with between 0 and 8 criteria, with a median of 2 and a mean of 2.6.
While 6 of the 95 studies did not specify any criterion (\texttt{n/a}), 22 studies provided 4 or more rationales.
As examples, \citeauthor{S7} did not provide any motivation why they used these snippets in S7, whereas \citeauthor{S84} used nearly a full page to describe their snippet rationales with admirable details in the dedicated subsection \enquote{Designing and Selecting Code Snippets} (S84, labeled with 8 criteria).
Snippet criteria fell into one of three categories: a) related to \textit{experiment design}, b) related to \textit{generalizability}, and c) related to \textit{participants}:

\begin{small}
\begin{table*}[t]
    \caption{Criteria for the creation or selection of experiment code snippets}
    \label{tab:snippetCriteria}
    \centering
    \begin{tabularx}{\textwidth}{p{4cm}rX}
        \toprule
        Snippet Criterion & \# of Studies & Description \\
        \midrule
        balance between simplicity and complexity & 42 & simple/small enough to be understood in a reasonable time, but complex/large enough to require some effort and to be realistic \\
        homogeneity concerning potential confounders & 29 & minimize the influence of unrelated factors such as naming styles, LoC, line length, formatting, or indentation \\
        novice friendliness & 25 & common beginner programming tasks, typical text book problems or algorithms known to students such as shell sort or binary search \\
        taken/adapted from existing studies & 21 & (partially) using/adapting snippets from existing studies against the results can be compared \\
        real-world code & 19 & taken/adapted from open-source or industry code \\
        self-contained functionality & 19 & no additional context or external functions necessary, e.g., a function, static method, main method, test class, etc. with a determinable purpose \\
        heterogeneity concerning the targeted constructs & 18 & differences in code metrics, control structures, method chains, comments, etc. to increase variance and generalizability \\
        requires bottom-up comprehension & 15 & obfuscate identifiers, remove beacons and comments, etc. to force line-by-line understanding \\
        thematic heterogeneity & 13 & a different algorithm, domain, etc. per snippet to avoid learning effects or to increase generalizability \\
        no unnecessary cognitive load & 12 & small inputs, simple arithmetic, no recursion, no non-deterministic or non-portable code, low computational complexity, no advanced OO, etc. \\
        no special domain knowledge necessary for understanding & 10 & avoid specific domains to reduce bias due to participant (in)experience, use familiar use cases like Pacman \\
        written in a popular/familiar programming language & 9 & mainstream language or familiar to participants, e.g., Java, C, Python \\
        n/a & 6 & not explicitly specified in the paper \\
        small enough to avoid extensive scrolling & 5 & code fits completely on screen, especially important for, e.g., fMRI or eye-tracking studies \\
        random sampling & 5 & snippets are randomly drafted from a pool \\
        no extensive usage of (rare) APIs & 4 & avoid specific APIs to reduce bias due to extensive language experience \\
        never seen before by participants & 3 & self-created/adapted to be new to participants \\
        code written by participants & 1 & participants needed to be familiar with the code and should have written it themselves\\
        \bottomrule
    \end{tabularx}
\end{table*}
\end{small}

\textit{a) Experiment design:} By far the most used criterion (42 of 95 studies) was that snippets needed to fulfill a delicate \textit{balance between simplicity and complexity}.
They had to be simple and small enough for participants to understand within a reasonable timeframe, yet large and complex enough to retain realism and require minimal effort.
What this meant in practice, however, could be very different depending on the studied constructs and the experiment design:
\citeauthor{S39} (S39) only used snippets between 3 and 24 LoC, \citeauthor{S5} (S5) aimed for between 30 and 70 LoC, while \citeauthor{S10} (S10) argued that between 50 and 100 LoC would be ideal for their experiment.
Another prominent dichotomy was that snippets should possess both \textit{homogeneity concerning potential confounders} (29) and \textit{heterogeneity concerning the targeted constructs} (18).
In this sense, researchers normalized unrelated factors in their snippets, e.g., naming styles or formatting, but created differences within their studied constructs such as code metric ranges or control structure usage to increase variance or generalizability.
A special case to avoid confounders was the conscious \textit{elimination of unnecessary cognitive load} (12).
Researchers removed or avoided, e.g., complex arithmetic, recursion, or high computational complexity to allow participants to focus on the essential comprehension aspects of the study.
In 19 studies, it was important for researchers that snippets represented \textit{self-contained functionality} with a determinable purpose.
These were, e.g., complete functions, static methods, or main methods.
Additionally, 15 studies ensured that a snippet \textit{requires bottom-up comprehension} by obfuscating identifiers and removing beacons and comments, thereby forcing line-by-line understanding.
Lastly, five studies ensured that a snippet was \textit{small enough to avoid extensive scrolling} because some fMRI or eye-tracking experiments required that snippets fit completely on screen.

\textit{b) Generalizability:} The most prominent criterion in this category was using snippets \textit{taken or adapted from existing studies} (21) to build on previous knowledge and to make results comparable.
Similarly, several papers required \textit{real-world code} (19), i.e., that snippets were taken or adapted from either open-source projects or proprietary industry projects to be realistic.
A few studies combined this with \textit{random sampling} (5) from a large corpus of open-source projects.
Furthermore, 13 studies prioritized the \textit{thematic heterogeneity} of snippets and used, e.g., different algorithms or application domains per snippet, mostly to increase generalizability, but sometimes also to avoid learning effects.

\textit{c) Participants:} Since a large portion of the studies relied on students as participants, a prominent snippet criterion was \textit{novice friendliness} (25).
Therefore, common programming tasks for beginners and typical problems or algorithms from text books were selected.
Similarly, some researchers also ensured that \textit{no special domain knowledge was necessary for understanding} (10) to not be limited to participants from certain backgrounds.
In some cases, popular use cases or applications were chosen, e.g., \citeauthor{S20} used the game Pacman in S20.
To increase their sampling pool, researchers also prioritized code \textit{written in a popular programming language} (9) that was familiar to their participants (e.g., Java, Python, or C) and \textit{avoided extensive usage of (rare) APIs} (4).
Lastly, three studies ensured that their snippets were \textit{never seen before by participants} and a single study (S42 by \citeauthor{S42}) required \textit{code written by participants} so that they were very familiar with it.

\subsubsection{Summary}
Code comprehension experiments have primarily been conducted onsite, although remote studies have emerged since 2007 and may be more prominent since 2019. Nowadays, code displayed on a screen is the rule, with the dominant programming languages utilized are Java, followed by C, C++, and Pascal. Researchers typically create code snippets themselves, but they also draw from open-source projects or previous studies. The criteria for selecting code snippets vary and are often not explicitly stated. However, when mentioned, researchers commonly consider factors such as achieving a balance between simplicity and complexity, controlling for potential confounders, and ensuring novice-friendly code.

\subsection{Comprehension Tasks and Measures}

The core of any code comprehension study, at least from the participant's point of view, is the actual code comprehension task.
Such a task, sometimes several in a single study, provides data for the analysis of a participant's code comprehension performance.
The precise nature of this analysis and the data collected for it are defined by comprehension measures.
In this section, we first look at the variety of designed comprehension tasks, then at the variety of comprehension measures, and finally at how comprehension tasks and comprehension measures co-occur.

\subsubsection{Comprehension Tasks}

Table~\ref{tab:comp_task} shows the variety of comprehension tasks and how many studies implemented which type of task.
The classification is a refinement of the division used by~\citet{Oliveira:2020:Evaluating} to group comprehension and reading tasks in their study.
All tasks could be assigned to four overarching task categories (Tc): provide information about the code (Tc1), provide personal opinion (Tc2), debug code (Tc3), and maintain code (Tc4).

\begin{table}[t]
\caption{Comprehension tasks that participants had to work on}
\label{tab:comp_task}
\begin{tabularx}{\columnwidth}{p{0.3cm} l X}
\toprule
& Comprehension Task & Studies \\
\midrule
\multicolumn{3}{l}{\textbf{\textit{Tc1: Provide information about the code}}}\vspace{0.1cm} \\
& Answer comprehension questions & S5, S8, S31, S32, S33, S34, S36, S37, S43, S44, S45, S51, S52, S54, S56, S62, S63, S67, S68, S69, S71, S76, S77, S80, S81, S82, S83, S86, S89, S90, S91, S92, S93\\
& Determine output of a program & S1, S2, S5, S7, S9, S11, S12, S13, S14, S15, S16, S17, S23, S25, S37, S39, S48, S50, S57, S59, S60, S68, S70, S77, S84, S88 \\
& Summarize code verbally or textually & S4, S11, S20, S24, S29, S35, S36, S37, S38, S42, S46, S47, S49, S58, S61, S72, S75, S77, S79, S87, S94, S95\\
& Recall (parts of) the code & S29, S44, S61, S66, S72, S73, S74 \\
& Determine code trace & S35 \\
& Match with diagram or similar code & S77, S84, S85\\\\

\multicolumn{3}{l}{\textbf{\textit{Tc2: Provide personal opinion}}}\vspace{0.1cm} \\
& Rate code comprehensibility &  S4, S6, S34, S36, S42, S49, S53, S69, S70, S75 \\
& Rate own code understanding & S5, S31 \\
& Rate task difficulty or task load & S12, S14, S23, S38, S51\\
& \makecell{Rate confidence in answer/\\understanding} & S23, S24, S33, S94\\
& Compare or rank code snippets & S21, S22, S40, S78\\
& Other subjective indications & S21, S22, S33, S41, S46, S78, S85\\\\

\multicolumn{3}{l}{\textbf{\textit{Tc3: Debug code}}}\vspace{0.1cm} \\
& Find a bug & S3, S10, S28, S38, S55, S64, S87 \\
& Fix a bug & S26, S27, S77\\
& Determine if code is correct & S33\\\\

\multicolumn{3}{l}{\textbf{\textit{Tc4: Maintain code}}}\vspace{0.1cm} \\
& \textit{Cloze test} on code & S4, S30, S74\\
& Extend the code & S21, S26, S27\\
& Refactor code & S65, S82\\
\bottomrule
\end{tabularx}
\end{table}

We found 81\% of all papers (77/95) to use at least one task that requires participants to provide information about a code snippet to be understood (Tc1).
Among these, 67.5\% exclusively used tasks from Tc1.
Three tasks in particular stand out as most frequently used: answering comprehension questions (33), determining the output of a program (26), and summarizing code verbally or textually (22).

A total of 27 (28.4\%) of the papers were interested in a personal assessment of the participants (Tc2), of which 7 exclusively used tasks of this category.
Preferably queried was a rating for the perceived code comprehensibility, in two cases for one's own code comprehension, four times for the confidence in one's own comprehension, in five papers the difficulty of the task was to be assessed, and in four papers code snippets were compared or ranked regarding the comprehensibility of other code snippets in these studies.
In 7 studies, subjective evaluations were requested, which differ from all others and are therefore listed under \enquote{other subjective indications} (e.g., S46: provide opinion of identifier type importance, or S78: identify most helpful code lines).

For 11 studies, a debug task (Tc3) was used to derive statements about code comprehension. Five of them exclusively considered this task category.
In seven papers, and thus most frequently in Tc3, participants had to find a bug.
In three studies, a bug had to be fixed. One study required a judgment about whether the code was correct or not.

We identified eight papers that asked participants to complete a maintenance task (Tc4).
In three cases, a Cloze test was used, a test in which gaps in the code must be filled.
In three studies, code had to be extended, and in two studies, code had to be refactored.
An example of the latter case is from \citeauthor{S82} (S82), who investigated the differences in the comprehension effectiveness between refactoring code and reading code.

While 66 of the 95 papers used exactly one task category, it was three for S4, S33 and S38.
In two studies (S18 and S19), we could not find any information about the comprehension tasks in the paper.
Perhaps the most unusual comprehension task was reported by \citeauthor{S90} (S90), who presented their comprehension questions as crimes that a detective had to solve based on information from an informant.
The peculiarity of this study was that the sample consisted of non-programmers, and therefore an interesting scenario had to be created.

\subsubsection{Comprehension Measures}

Almost all studies have measured how well participants perform in comprehension tasks. The three studies S16, S17 and S54 were the exception; participants had to understand code, but there was no attempt to quantify how well they understood it, since, for example, the focus was on the observation of psycho-physiological responses only.
In all other cases, the comprehension measures used can be divided into six categories: correctness, time, subjective rating, physiological, aggregation of the former, and others.

\begin{figure}[b]
    \center
    \includegraphics[width=\columnwidth]{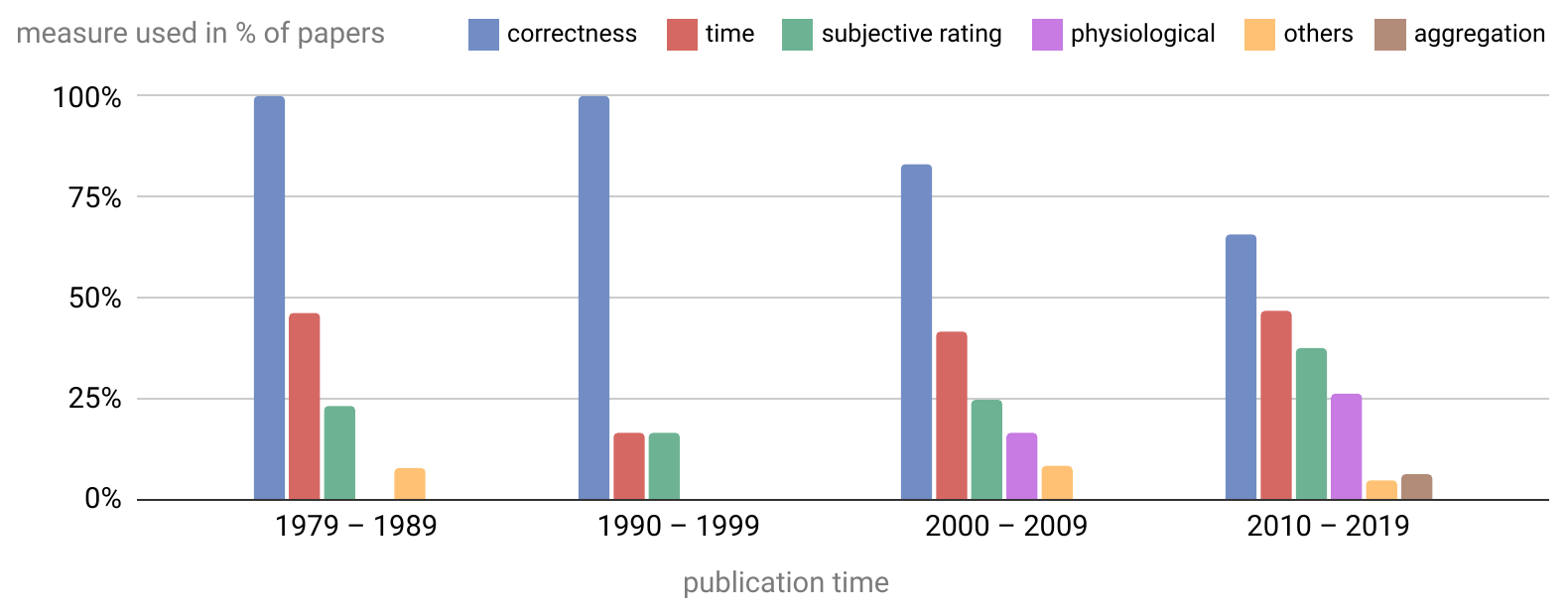}
    \caption{Comprehension measures used relative to number of papers}
    \label{fig:measuresRelPapers}
\end{figure}

Fig.~\ref{fig:measuresRelPapers} shows the percentage of comprehension measures to the number of papers in each period.
In all decades, the share of papers that measured at least correctness was highest.
Before 2000, correctness was even measured in all published papers, but the share decreased to 65.6\% in 2010--2019.
The proportion of papers that asked for subjective self-assessments or measured time increased over the years.
Most notable, however, is the increase in the use of physiological measures, which were used in over a quarter of publications in 2010--2019.
Furthermore, the diversity of measures used has never been greater than it is today.

Of the 95 papers, only 37 measure code comprehension in a single way. 36 papers rely on 2, 15 papers rely on 3, and 4 papers rely on 4 measures from different categories.
The measurement of correctness and time was the most frequent combination of measures with 23 papers, if more than one measure was collected.
In the period 2010--2019, we also identified for the first time four papers that not only measured multiple comprehension measures in their study, but even aggregated them.
This involved combining correctness and time in three studies (S9, S26, S27).
\citeauthor{S5} (S5) aggregated correctness, time, and subjective ratings in different ways: by combining correctness and time, but also correctness and subjective ratings, and time and subjective ratings.

At this level of granularity, it is possible to summarize very well which trends there have been over the years and which way of measuring code comprehension seems to be the most popular.
It is also closest to the terminology used in the primary studies themselves for the various measures if an abstraction of the specific calculation is used in the report (e.g., \enquote{The metrics used to investigate these questions are time and correctness [\ldots]}~\cite{Ajami:2018:Syntax}).
A closer look, however, reveals that almost every paper differs in the concrete implementation of the measures.

Consider \textit{correctness} as an example.
With 70 papers (73.7\%), it is the most prominent comprehension measure in our dataset.
Our primary studies often measured it via the relative or absolute number of correct answers to comprehension questions or bug finding tasks.
However, we also found the usage of f-measures (e.g., S63, S65), number of correctly recalled identifiers or statements (e.g., S29, S73), number of fails until a correct answer was given (e.g., S28), number and severity of errors (e.g., S48), as binary variable (e.g., S61, S62), manually rated functional equivalence of recalled code (e.g., S73), automatically or manually rated code changes (e.g., S20, S26), and manually rated free-text or oral answers (e.g., S24, S29, S36, S39, S79, S94).

For the measurement of time and subjective ratings, there is a similarly large variance in the concrete implementation.
For more detailed insights, we refer the interested reader to our dataset.
At this point, we would rather elaborate on the measurement of code comprehension via physiological measures, since these are still relatively novel in the field of code comprehension as a whole.

The category of physiological (or psycho-physiological) measures includes the measurement of code comprehension via responses of the human body or brain to performing code comprehension tasks.
In our dataset, we encountered the measurement of fMRI data (brain activation regions, concentration levels, brain activation strength, cognitive load, blood oxygen level dependent), fNIRS data (regional brain activity), and eye-tracking data (visual/mental effort, e.g., on gaze behaviors such as number of fixations and their duration).

The primary studies that use such measures already provide initial findings on physiological responses in code comprehension, but are currently still largely to be understood as feasibility studies.
For example, they investigate correlations with conventional comprehension measures (see, e.g., S1, S12--S19 and also Section~\ref{subsec:studythemes}).
In the medium term, psycho-physiological measures could provide objective information about the process and performance of an individual during code comprehension~\cite{Fakhoury:2018:ObjectiveMeasures} and, e.g., help to investigate \enquote{the role of specific cognitive subprocesses, such as attention, memory, or language processing}~\cite{Peitek:2018:Simultaneous}.

Finally, we note that we could not assign measures to the categories described above in only five cases.
These measures ended up in the \enquote{other} category.
S21 and S59 used code metrics to measure code comprehension, S28 visual focus via a sliding window (not via physiological measures), S47 the number of animation runs of a code visualization tool, and S81 the number of times an algorithm was brought into view.
However, none of the five corresponding papers exclusively used these measures, so they are also represented in at least one of the other measure categories.

\subsubsection{Combinations of tasks and measures}
Counterintuitively, comprehension tasks and comprehension measures were rarely mutually dependent in their selection.
Almost all comprehension measures, or at least the categories presented, can be applied to all types of comprehension tasks that we identified in this mapping study.
Nevertheless, some combinations occur significantly more often than others.
The combination of correctness and Tc1 (provide information about the code) appears most frequently; about two thirds (66.3\%) of all papers use at least this combination of measure and task, and 16.8\% of all papers use this combination exclusively.
If debugging tasks (Tc4) are used, then at least correctness is measured in all cases.
If, in contrast, tasks from Tc2 (provide personal opinion) are used, correctness only occurs in 55.6\% of these cases in the paper.
Tc2 tasks are mainly evaluated by subjective ratings (96.3\% of all tasks in this task category appeared in papers at least with subjective rating measures).

\subsubsection{Summary}
We observe a shift in the use of measures over time, both in terms of an increasing variety and a decreasing dominance of individual measures.
While correctness continues to be the preferred measure of code comprehension, the use of psycho-physiological measures is increasing rapidly.
Further, the concrete design of comprehension tasks is in itself very diverse.
Most studies use tasks that ask for information about the code or report subjective assessments of code understanding by the study participants themselves.
On top of this variety of measures and tasks, there are almost unlimited was to combine tasks and measures, making almost every paper unique in its way of measuring how well a participant understood source code.
We discuss the consequences of this observation in Section~\ref{sec:discussion}.

\subsection{Reported Limitations}
\label{reported_limitations}

We extracted study design limitations described by the authors of the primary studies.
Already in the extraction of the limitations, we have only extracted those limitations that the authors themselves also considered as such and not aspects that the authors explicitly assume not to be threats because, for example, they have been mitigated.
The extracted threats were then analyzed qualitatively and quantitatively in several steps, which we describe below.

Each limitation was labeled, and the resulting label was assigned to a class of validity threats.
The classification is one that is usually used for reporting positivist empirical
studies in software engineering~\cite{Wohlin2012,Jedlitschka:2008:Reporting}: internal, external, construct, conclusion, and other validity threats.
If the authors themselves have made use of this classification, we have followed their distinctions (which was the case for 30 papers).
Otherwise, we have classified the labels ourselves by following the descriptions provided by~\citet{Jedlitschka:2008:Reporting}:

\begin{itemize}
    \item \enquote{Internal validity refers to the extent to which the treatment or independent variable(s) were actually responsible for the effects seen to the dependent variable.}
    \item \enquote{External validity refers to the degree to which the findings of the study can be generalized to other participant populations or settings.} 
    \item \enquote{Construct validity refers to the degree to which the operationalization of the measures in a study actually represents the constructs in the real world.}
    \item \enquote{Conclusion validity refers to whether the conclusions reached in a study are correct.}
    \item \enquote{Other threats than those listed above may also need to be discussed, such as personal vested interests or ethical issues regarding the selection of participants (in particular, experimenter-subject dependencies).}
\end{itemize}

\noindent The individual labels are similar in part because different studies report similar limitations.
We therefore categorized the labels to better grasp the diversity of limitations.
The work by~\citet{Siegmund:2015:Confounding}, who compiled a list of confounding parameters in program comprehension studies, builds the basis for these categories.

Since~\citeauthor{Siegmund:2015:Confounding} focused on confounders, their categories and descriptions are mainly focused on the consequences for internal (and construct) validity.
We nevertheless tried to stay close to this list to allow for comparability of results.
To accomplish this, we have slightly renamed a few categories and added missing categories as needed.
Existing category descriptions were rewritten so that they answered the question of how the authors of a primary study think that an aspect of validity of their study could be limited or threatened.
All category descriptions are part of our supplemental materials.

The specific aim of this review of reported limitations is to assist researchers in designing new studies by providing them with a list of threats that other researchers have previously reflected on for their design.
We therefore considered it useful to group the categories according to different design aspects.
For example, when selecting code snippets, one can specifically look at the list of threats found under \enquote{code snippet selection}, and when designing a comprehension task, consider the threats under \enquote{task design}.
In such a phase of the study design, all potential threats have to be assessed anyway, regardless of how their consequences can be classified.
Therefore, threats to internal, external, construct, and conclusion validity can be equally represented behind the respective categories.

\subsubsection*{Results}

Of the 95 studies, 79 reported threats to validity.
Interestingly, for the other 16 (16.8\%), no correlation with publication year or venue can be found.
Nine papers were published up to 2006, so they tend to be older.
However, the remaining seven papers without threats to validity were published since 2017, i.e., in the past five years.
Moreover, 13 of them are either journal or conference articles.
Only three are workshop papers, where one can assume that space limitations are the reason for not reporting limitations.

\begin{figure}[h]
    \center
    \includegraphics[width=0.7\columnwidth]{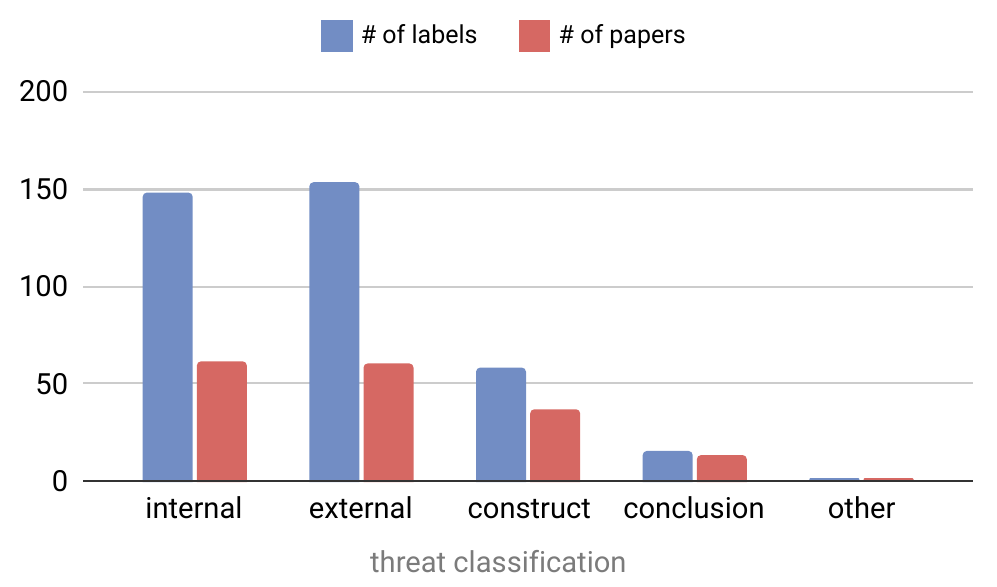}
    \caption{Number of labels and papers within the threat classification}
    \label{fig:labelCount}
\end{figure}

The analysis resulted in 376 labels, which is equivalent to 376 individual, non-unique threats to validity that have been reported.
As can be seen in Fig.~\ref{fig:labelCount}, most of the labels are equally distributed between internal (39.4\%) and external (40.7\%), followed by construct validity threats (15.4\%) and rare cases of conclusion validity threats (4.3\%).
One can assume that at least one internal or one external threat is reported in a paper if threats to validity are reported.
Nevertheless, at least one construct validity threat is reported in 46.8\% and at least one conclusion validity threat in 17.7\% of the papers that report any threats to validity.

The only label we assigned to the \textit{other} class was a threat from study S17, which can be described as a high data drop-out due to the chosen comprehension measures (fMRI and eye tracking) in a proposed methodology.
Since this class of \textit{other} threats did not occur significantly often, we will only discuss internal, external, construct, and conclusion threats in further reporting for simplicity.

Table~\ref{tab:reported_limitations} constitutes the main contribution of this section.
It lists 52 threat categories, 33 of which come from the study by~\citet{Siegmund:2015:Confounding}. 
For each category, the number of assigned labels and the distribution of the labels among the threat classifications are provided.
In the following, we highlight some interesting findings.

\begin{scriptsize}
\input{_table_reported_limitations}
\end{scriptsize}

First, we did not assign labels to three categories that we adopted from~\citet{Siegmund:2015:Confounding}: \textit{Color blindness}, \textit{Reading time}, and \textit{Treatment preference}.
This may be explained by our restriction to bottom-up source code comprehension studies, which means that we only consider a subset of the program comprehension papers that~\citet{Siegmund:2015:Confounding} analyzed and that these threats happened to be discussed only in the other papers.
Threats in these three categories also occurred comparatively rarely in the analysis by~\citet{Siegmund:2015:Confounding}.
We nevertheless left these categories in the table for ease of reference.

Second, many discussed threats are reported quite generically in our primary studies, i.e., without a clear description of what the actual threat is and without an explanation, why a certain design property is a limitation.
For example, with 43 label occurrences, the rather generic category \textit{Sampling strategy} is the largest threat category in our dataset.
It contains, e.g., descriptive statements of the resulting participant sample of a primary study, such as all participants being students, and that it is unclear whether such a sample can be generalized to industrial professionals.
Labels in this category do not provide a rationale for why such a threat would exist, although we identified 17 other participant-related categories that would allow for more nuanced assumptions.
Speaking of assumptions, it is furthermore striking that hardly any studies explicitly cite any evidence at all for the suspected threats to validity.
We address the issue of potential evidence gaps and the need for more intensive reflection on the threats to one's own study in more detail in our discussion.

Third, in some cases, threat classes dominate individual aspects of a study design.
For example, the selection and design of the \textit{materials} predominantly has a presumed influence on the external validity of a study.
Researchers often discuss that their code snippets are relatively short (\textit{size of study object}) and therefore the generalizability to longer snippets is unclear.
By contrast, when it comes to \textit{research design, procedure and conduct}, consequences for internal validity are discussed almost exclusively: \textit{learning effects} and general threats to ensure causal relationships (\textit{causal model}) are the categories with the most labels here.

At the same time, these are examples of threat categories that mutually affect each other.
More extensive code snippets would lead to more variance in the snippet features, and thus to more potentially uncontrolled influences that make causal inferences difficult.
The researcher must now balance between strengthening internal validity and strengthening external validity.

Fourth and finally, measuring how well someone has understood code is apparently difficult.
It is not only the case that researchers in code comprehension studies have to deal with potential threats to validity from over 50 categories.
In the end, 39 labels were assigned to the \textit{instrumentation} category and another 9 to the \textit{construct operationalization} category, two categories that both directly address the fundamental question of how the performance or process of code comprehension is operationalized and measured.
In the respective papers (e.g., S4, S9, S63, S69), it is sometimes discussed quite openly that there is uncertainty about whether the chosen way to operationalize the construct of understanding, e.g., via time and correctness, is adequate at all, and that the used measurement instruments have validity issues.
Such insights point to the need for the research community to achieve more certainty in the design of code comprehension studies through discourse and research in the future, and strengthen our motivation for this work to take a step in this direction with a synthesis of the state-of-the-art.

%
% -------------------------------------------------------------------------- %
% Discussion
% -------------------------------------------------------------------------- %
%

\section{Discussion}
\label{sec:discussion}

One of the motivations for our systematic mapping study was to provide researchers with confidence in their study designs by highlighting what is considered acceptable by the research community.
This should not only serve as a helpful overview for newcomers to the field, but may also enrich the perspective of seasoned researchers on the topic, e.g., by providing insights into trends and shortcomings.
To summarize which design decisions have dominated the research field in the past, we therefore start this section with a description of the typical code comprehension experiment.

\subsection*{Answer to RQ2: What are differences and similarities in the design characteristics of code comprehension experiments?}

On the one hand, there \textit{are} certain design options that are taken more frequently than others.
For example, the majority of studies examine the impact of a construct on code comprehension, using one-at-a-time within-subject designs.
There are usually fewer than 50 participants sampled, of which at least some are students.
The used code snippets are often specifically created for the study, Java is clearly the programming language of choice, and participants are shown the snippets on a screen.
When it comes to understanding the snippets, participants are required to provide information about the code, e.g., by answering comprehension questions or determining the output for a given input.
The typical evaluation assumes that the faster participants provide the correct answer, the better their understanding.
Last but not least, researchers are aware that every study design comes with limitations, which is why almost all papers discuss threats to validity.

On the other hand, it also became clear that \enquote{the typical code comprehension experiment} does not exist.
Each study in our dataset is unique.
For example, there is an incredible variance in the design of the concrete comprehension tasks and measures.
Almost all studies come up with their own individual task design, even though all of them \enquote{only} try to measure how well a participant understood a certain code snippet.
Why does every study develop its own code comprehension tasks?
Are existing study designs not convincing, or does every research question actually require an individual approach?

We suspect that part of the variance is due to uncertainty.
The most important driver in the design of code comprehension studies is currently the intuition of the researchers behind the studies.
There is a lack of theoretical foundations on which all code comprehension studies could be built.
Such foundations should offer (1) a common definition of code comprehension, (2) a mental model describing how code is represented in a developer's mind and (3) a cognitive model that explains the process of code comprehension.
A strong theory with these three parts would greatly help with anchoring new research designs and integrating empirical findings.
Without such foundations, you cannot help but rethink almost every design decision to answer a research question, as it is simply not apparent from existing primary studies whether assumptions made and views held about an underlying code comprehension model are consistent with your own.
In other words, so far, everyone has been doing it in a way they think is reasonable --- after all, no one \textit{really} knows what is reasonable at the moment.

In the 1980s, several cognitive models have been proposed for code comprehension.
Important contributions were the bottom-up comprehension strategy that describes the chunking of parts of the code to higher-level abstractions~\cite{shneiderman1980software}, the differentiation of syntactic and semantic knowledge~\cite{shneiderman1979syntactic}, and types of mental models developed during comprehension such as a program model and a situation model~\cite{pennington1987stimulus}.
In contrast, \citeauthor{Brooks:1983:TowardsTheory}~\cite{Brooks:1983:TowardsTheory} proposed that developers comprehend code in a top-down manner by forming hypotheses and using beacons to verify these hypotheses.
\citeauthor{Soloway:1984:Empirical}~\cite{Soloway:1984:Empirical} added the concepts of programming plans that represent typical, generic programming scenarios and rules of discourse describing coding conventions.
\citeauthor{vonmayrhauser1995}~\cite{vonmayrhauser1995} integrated these and other theories into their metamodel of program comprehension in the 1990s.
Since then, only very few theories have been proposed, with little impact on the comprehension research community (e.g., \cite{murray2005generating,gueheneuc2009theory}).

We expected that these models, in particular the integrated model~\cite{vonmayrhauser1995}, would be the basis to inform hypothesis development and experiment design for code comprehension studies.
Yet, in most experiments, this seems not to be the case.
Regarding the \emph{hypotheses} and \emph{study purposes} of the primary studies, we partly found concepts of comprehension theories, although often without explicitly stating this connection.
Semantic cues and mental models are important concepts in the integrated model.
Likewise, code structure and visual characteristics are relevant for the theory of program model structures.
Yet again, these concepts were most often not directly related to the respective theory, and hence, do not act as rationales for the chosen design.
The major exception is S84, a study by \citeauthor{S84}, who explicitly used the concepts \textit{bottom-up comprehension}, \textit{semantic comprehension}, and \textit{programming plans} to motivate their experiment design.

The \emph{study themes} we identified in the primary studies are also only partly related to comprehension theories.
Examples of theoretical concepts are \textit{beacons}, while \textit{identifier naming} or \textit{control structures} are at least related to theoretical concepts.
Similarly, the often used \textit{experience} is related to program-domain knowledge in the integrated model.
Concerning the \emph{code snippets} used in the studies, grounding their properties in theoretical concepts seems especially important for an empirical investigation.
For example, the integrated model assumes rules of discourse about how code is usually written, and that deviating from these rules will make comprehension more difficult.
Hence, we would expect an evaluation of this characteristic for the used snippets.
However, we rather saw discussions whether open-source code or artificially generated code is used.
This is related, but not the same concept.
As a positive example, there was usually a clear reference to theory when researchers obfuscated identifiers or removed beacons to force bottom-up comprehension.

In our opinion, the most pressing issue in this area is that the selection of \emph{tasks} and \emph{measures} is not based on the integrated (or any other) theory.
Measures like the correctness of answers to comprehension questions are somewhat related to the mental model of the participant.
Yet, in the integrated model, we have top-down structures, program model structures, and situation model structures.
Are the questions covering all three of them, or only one?
How does this impact the results of the study?
A study could, e.g., explicitly aim for understanding the impact of certain semantic cues in the code only on the situation model.
We found only~\cite{Shneiderman:1977:Measuring} evaluating explicitly the developer behavior theory proposed in~\cite{shneiderman1979syntactic} by using recall measures.
All other studies seem not to rely on a theoretical foundation for their tasks and measures.

Again, we consider diversity in study designs to be important for addressing research questions from multiple perspectives.
However, each study design in itself should be based on logically sound design decisions.
This is not currently possible, or at least difficult to evaluate, as basic research on code comprehension has diminished significantly since the 2000s. 
The earlier drivers of such research, e.g., on the distinction between bottom-up and top-down comprehension, have not progressed to the point where we could already support entire study designs with their models.
We will come back to this when we conclude our paper in Section~\ref{sec:conclusion} with specific calls to action to the research community.
Before we do so, however, we will discuss a few other peculiarities that we noticed when reviewing code comprehension experiments of the past forty years.
Note that the following points all draw on our empirical findings, but we deliberately refrain from citing specific negative examples because we do not consider this necessary.
Instead, we highlight positive examples where available.

\subsection*{Answer to RQ3: What are differences and similarities in the reporting characteristics of code comprehension experiments?}

While we already described the existing heterogeneity of study designs as surprising, the variety of ways in which these designs have been reported in the papers appeared even more substantial to us.
A few authors tried to strive for a more consistent reporting via tables with the most important characteristics, by following the experiment reporting guidelines by \citet{Jedlitschka:2008:Reporting}, or by reporting threats to validity according to the categories by \citet{Wohlin2012}.
However, as partially illustrated by the results in Section~\ref{sec:results}, there were strong differences between papers regarding where and how the same aspects of the study designs were reported, and especially if rationales for certain design decisions were provided.
This did not only concern the used level of detail and the location in the paper, but also if something was reported \textit{at all}.
Typical aspects of the study design that were reported very differently were, e.g., the studied construct and its definition, participant characteristics, sampling of participants, experiment setting, and threats to validity.
We provide more concrete insights into issues with these reporting differences below when discussing RQ4 and our proposed action items in Section~\ref{sec:conclusion}.

\subsection*{Answer to RQ4: Which issues and opportunities for improvement are evident in the design and reporting of code comprehension experiments?}

\paragraph{Lack of studies investigating the impact of code comprehension on other constructs}
In our sample, not a single study investigated the effects of understanding code on something else.
During study selection, we also did not find such studies that we had to exclude, e.g., due to a lack of human participants or a focus on top-down comprehension.
Most may agree that high code comprehensibility is a desirable goal in itself, but, for others, it may become even more tangible and important if consequences of good or bad code comprehensibility were empirically supported.
Examples are how code comprehension impacts the motivation to work on the code,  general job satisfaction, or accuracy in the effort estimation for code modifications.
This type of experiment seems extraordinarily rare, even though some studies exist that analyze this for higher-level constructs that (partly) include code comprehension, e.g., the impact of technical debt on developer morale~\cite{Besker2020}.

\paragraph{No definitions for the studied construct(s)}
For the majority of our primary studies, the authors do not provide a clear definition or description for the central construct(s) that they want to measure.
Rare shining examples are S1, where \citeauthor{S1} clearly describe the differences between bottom-up and top-down comprehension, and S5, where \citeauthor{S5} do the same for understandability and readability.
As a consequence, some authors call their construct \enquote{readability}, despite clearly measuring comprehension.
Likewise, several studies use the older and more general term \enquote{program comprehension}, even though they measure bottom-up source code comprehension. That the \emph{program} comprehension research field today encompasses much more than comprehension of source code can be seen, for example, in the small number of papers in our dataset that were published at the ICPC. Therefore, it would be much more explicit if researchers used the term code comprehension instead of program comprehension whenever it applies.\\
This state of inconsistency in naming and defining the construct of interest makes it very hard for primary studies to compare their findings with those of others.
It further severely increases the required effort for secondary studies. We spent countless hours figuring out whether studies were actually measuring bottom-up comprehension.

\paragraph{No clear operationalization for the construct}
In several studies, the researchers did not explicitly state how they collected a measure like \enquote{correctness}.
Sometimes, they also collected multiple measurements, but did not explicitly state which of those measurements are used to operationalize the construct during the analysis.
For a study that measures, e.g., the time to read code and the time to answer comprehension questions, it must become clear if only one of the measures or a combination of both was used in the analysis.
A good example is again provided in S5, where \citeauthor{S5} clearly describe their aggregations of correctness, time, and subjective ratings to evaluate their construct.

\paragraph{Unsuitable comprehension tasks}
While there is no consensus (yet) on optimal tasks for code comprehension experiments, there are definitely some tasks that seem less suitable than others.
Several studies used a recall task, where participants had to recite the (exact) line-by-line code snippet to judge their comprehension.
Comprehending means to form a mental structure that represents the meaning of code.
From research on text comprehension, we know that \enquote{whatever mental structure is incidentally generated in the process of comprehension also serves as a retrieval structure}~\cite{Kintsch:1998:Comprehension}.
Shneiderman similarly hypothesizes that developers who perform better on a recall task have a better understanding of the program~\cite{Shneiderman:1977:Measuring}.
However, this is not a sufficient criterion for assessing comprehension.
An evaluation with recall tasks must be designed very carefully because being able to recite something flawlessly from memory is not the same as truly comprehending its meaning or purpose.
Likewise, participants unable to remember the exact lines of code may still have understood what the snippet accomplishes. 
In this sense, we share \citeauthor{Feitelson:2021:Considerations}'s sentiment that \enquote{failure in recalling code verbatim from memory may identify totally wrong code or code that does not abide by conventions, not necessarily hard to understand code} and that such a task is far removed from what developers do in their daily work~\cite{Feitelson:2021:Considerations}.\\
In addition, we also found multiple studies that used tasks that do not measure comprehension in isolation but include much more, e.g., fixing a bug, refactoring code, or extending code.
This type of problem-solving differs from mere comprehension in the way that problem-solving is a \enquote{controlled, resource-demanding process involving the construction of problem spaces and specialized search strategies}, whereas comprehension can be simplistically summarized as an automatic process~\cite{Kintsch:1998:Comprehension}.
In the extreme case of S73, \citeauthor{S73} even let their participants write functionally equivalent code for a snippet.
While these tasks certainly require a certain degree of comprehension, they also introduce many more confounders and potential biases, and would be more suitable if the studied constructs were \textit{maintenance} or \textit{refactoring}.

\paragraph{Lack of closed-source code snippets from industry}
Only 3 of the 95 primary studies used code snippets from proprietary industry projects.
While open-source or constructed snippets may provide sufficient realism or industry relevance for many bottom-up code comprehension results, it is still noteworthy that basically no findings in this field are validated with closed-source industry code and that generalization is assumed.
Today, many companies are involved in some projects in open-source repositories.
Still, given the known differences between proprietary industry and open-source projects~\cite{Paulson2004,Robinson2010}, it would be highly unlikely that \textit{all} findings are fully transferrable.
Even if it seems plausible for the majority of scenarios, we simply do not know the extent.
We are aware that it is difficult to obtain closed-source industry code that can also be used with minimal modifications in experiments.
Additionally, this might also make sharing the code snippets with the publication impossible.
For now, though, it would at least be desirable to find out more about the influence of the differences between proprietary industry and open-source snippets on code comprehension and thus contribute to a more informed discussion about generalizability.

\paragraph{Incomplete reporting of experiment design characteristics}
Several studies omit important details of their experiments in the reporting, sometimes information so basic or essential that its absence was very surprising to us.
For example, we could several times  not reliably infer whether snippets were self-made or taken from somewhere else (14 papers), which programming language was used (7), whether the study was conducted remote or onsite (7), whether snippets were shown on screen or paper (6), why the respective snippets had been chosen (6), what type of participants were used (4), or how many snippets were used in total and per participant (4).
Many of these details are important to judge the soundness of the design or to interpret the results, e.g., concerning generalizability.

\paragraph{Inconsistent usage of categories for threats to validity}
While 79 of 95 studies included a description of threats to validity and many authors classified them into the typical categories, several instances of incorrect classifications made our aggregation more difficult.
For example, \citeauthor{S63} assigned potential confounding factors to construct validity instead of internal validity (S63), and \citeauthor{S68} associated the lacking representativeness of their code snippets with internal validity instead of external validity (S68).
This may be due to different definitions of the validity categories, but since most authors did not provide appropriate explanations or a reference for the used classification, we cannot say for sure.
At least, the inconsistent and partly incorrect categorization of threats to validity does not seem to be an issue exclusive to code comprehension studies (see, e.g.,~\cite{Ampatzoglou:2019:ValiditySecondary}).

\paragraph{Implicit or shallow reporting of threats to validity}
Many studies report both threats that were consciously mitigated and threats that remain.
This observation is in line with that of another literature study by~\citet{Sjoberg:2022:Construct} on reporting construct validity threats in software engineering.
Sometimes, however, it was difficult to distinguish between these two types because the authors did not explicitly mark the threats accordingly.
Therefore, interpreting the limitations of such studies becomes difficult.
\citeauthor{S3} present a good solution to avoid this issue in S3: they report how they consciously addressed common threats in their study design section, while the remaining threats are discussed separately at the end.
Moreover, a few papers reported very generic threats that  give the impression of simply pre-empting criticism from readers, or initially from reviewers of the manuscript.
For example, if the sample consists exclusively of students, it seems almost obligatory to mention this as a limitation (see Section~\ref{reported_limitations}).
However, it is rarely explained in the specific study context which student characteristics could actually threaten validity regarding the research questions.
Authors of primary studies that predominantly sampled students often cast doubt on the generalizability of their findings to the population of professional software developers, but do not explain why.
That such generalizability should not be questioned per se is shown at least by the existence of controversial discussions concerning student samples~\cite{Feldt:2018:FourCommentaries}.
How it could be done instead is shown by the authors of S26 and S82, who discuss, e.g., that the students in their sample may be influenced by the teaching assistant's coding style requirements or may not be representative because of their particular interest in the course.

\smallskip
\noindent
To counteract some of these identified shortcomings, we propose several calls to action for the research community in the final section below.

%
% -------------------------------------------------------------------------- %
% Conclusion
% -------------------------------------------------------------------------- %
%

\section{Conclusion}
\label{sec:conclusion}

We conducted a systematic mapping study to structure the research from 95 primary studies on source code comprehension published between 1979 and 2019.
Through this review of the past 40 years, we arrived at some interesting findings that should motivate changes for the future.
Below, we have focused on what we believe are the five most important action items that we should address as a code comprehension research community moving forward.

\begin{enumerate}
  \item \textbf{Work on a definition for code comprehension and provide such a definition in every primary study.} Code comprehension is a complex construct, making it all the more important to define what you intend to measure.
  While there are a few program comprehension models and a few vague definitions of what it means to understand a program, the landscape of definitions for code comprehension is unfortunately quite limited.
  A definition of code comprehension must not only be clear in describing the construct itself.
  It should also clearly distinguish code comprehension from other related constructs (see Section~\ref{sec:comprehension}).
  Without such a reusable definition, we burden each primary study with the task of defining and separating code comprehension from other constructs, which ultimately leads to a situation in which hardly any primary study defines what it actually intends to measure. Consequently, speculating based on the properties of the comprehension tasks is often the only way to assess whether two experiments intended to study the same construct.
  \item \textbf{Research the cognitive processes of code comprehension.} In short, we need more theory. Basic research on the cognitive processes involved in code comprehension, e.g., to discover different comprehension strategies and involved cognitive processes in the brain, is necessary to theoretically motivate a study design. Almost none of our primary studies linked their design to a comprehension theory that would justify, e.g., the used tasks and measures. Basic research on program comprehension is currently gaining momentum; the use of neuro-physiological measures to \enquote{develop a neuroscientific foundation of program comprehension}\footnote{https://www.se.cs.uni-saarland.de/projects/BrainsOnCode/} is particularly promising.
  \item \textbf{Conduct research on the comparability of different design characteristics.} We have seen that each of our 95 primary studies used a unique design to answer its research questions. We lack evidence on the consequences of particular design decisions, such as the choice of particular comprehension measures or experimental materials. We can only meaningfully compare primary studies and synthesize results in meta-studies if we know whether and how different design decisions affect study outcomes differently.
  \item \textbf{Gather evidence of commonly assumed consequences of discussed threats to validity.} Related to the previous point, we also lack evidence in discussing limitations of study designs. The catalog of potential threats to validity is too long to discuss in a primary study. Accordingly, the focus should be on those threats to validity that actually affect the results for a specific study design. However, to assess for which threats this is the case, we need more research on the influence of assumed confounding factors and generalization issues. Existing literature supporting the influence of specific design decisions on study results should then be referenced appropriately in the discussion of primary studies on code comprehension.
  \item \textbf{Agree on a set of design characteristics that must not be omitted when reporting a primary study.} Finally, we have a suggestion for improving the future reporting of code comprehension studies. As much as we desire to grant everyone their creative freedom in the organization of their papers, we nevertheless note that it was very cumbersome to extract the relevant design characteristics from some primary studies. Although we read all the papers from start to end, the relevant information was scattered throughout the report, sometimes ambiguous, not always where one would expect it to be, and sometimes missing completely.
  We would like to highlight S3 and S27 as positive examples that present a tabular overview of the \enquote{main factors} of the conducted study. This form of concise presentation is not only interesting for those conducting meta-studies~\cite{Wohlin:2014:writing}, but in our opinion also helps every reader to better understand the study design. As a community, we can discuss which specific design characteristics \textit{need} to be included in such table. For a start, the primary studies mentioned above as well as most categories of our Table~\ref{tab:data_items} provide a good starting point. 
\end{enumerate}

The research field is currently flourishing in terms of the number of new publications and number of researchers involved.
We are personally pleased to see this because we consider it an important field of research.
This makes it all the more important for us to emphasize at this point that, in our opinion, much of the design and reporting of code comprehension experiments is already of good quality.
The emphasis in the final pages of this paper on what we could improve in the future is not intended to slow down the growth in this field.
On the contrary, the five concrete action items are meant to encourage the community to use the momentum and engage even more deeply with code comprehension studies at the meta-level.

Since parts of the field are already changing, e.g., in the aspect that psycho-physiological measures are increasingly used in code comprehension studies, we consider an extension of our study to be interesting already in a few years.
At best, some of the more recent publications might then already reflect the influence of some of the proposed action items.

\bibliographystyle{ACM-Reference-Format}
\bibliography{main}

% \bibliographystyleS{ACM-Reference-Format}
\bibliographystyleS{unsrtnat}

\newpage
\bibliographyS{primary-studies}

\end{document}

%% file: _table_reported_limitations.tex
\begin{table*}[th]
\caption{Frequency of reported threat categories. For each category, a stacked bar chart indicates the share of threats assigned to one of the four classes (\textcolor{colorA}{internal}, \textcolor{colorB}{external}, \textcolor{colorE}{construct}, \textcolor{colorF}{conclusion})}
\label{tab:reported_limitations}
\begin{tabularx}{\textwidth}{l l l l p{10mm} X l l l p{10mm}}
\toprule
\multicolumn{3}{l}{\textbf{Threats Related To}} & \textbf{\#Labels} &  &  & \multicolumn{2}{l}{\textbf{Threats Related To}} & \textbf{\#Labels} & \\
\midrule
\multicolumn{3}{l}{\textbf{Participants}} & & &                             & \multicolumn{2}{l}{\textbf{Task design}} & &\\
  & \multicolumn{2}{l}{Individual background} & & &                         & & Difficulty & 8 & % stacked bar chart for share of limitation categories
  \def\INTERNAL{24}%
  \def\EXTERNAL{38}%
  \def\CONSTRUCT{38}%
  \def\CONCLUSION{0}%
  \input{_limBar}%
\\
  & & Color blindness & 0 & &                                               & & Time limit & 2 & % stacked bar chart for share of limitation categories
  \def\INTERNAL{50}%
  \def\EXTERNAL{50}%
  \def\CONSTRUCT{0}%
  \def\CONCLUSION{0}%
  \input{_limBar}%
\\
  & & Culture & 2 & % stacked bar chart for share of limitation categories
  \def\INTERNAL{0}%
  \def\EXTERNAL{100}%
  \def\CONSTRUCT{0}%
  \def\CONCLUSION{0}%
  \input{_limBar}%
 &                                 & & Task procedure & 14 & % stacked bar chart for share of limitation categories
  \def\INTERNAL{21}%
  \def\EXTERNAL{72}%
  \def\CONSTRUCT{7}%
  \def\CONCLUSION{0}%
  \input{_limBar}%
\\
  & & Gender & 3 & % stacked bar chart for share of limitation categories
  \def\INTERNAL{0}%
  \def\EXTERNAL{100}%
  \def\CONSTRUCT{0}%
  \def\CONCLUSION{0}%
  \input{_limBar}%
 &                                  & & Task description / study introduction & 5 & % stacked bar chart for share of limitation categories
  \def\INTERNAL{20}%
  \def\EXTERNAL{20}%
  \def\CONSTRUCT{60}%
  \def\CONCLUSION{0}%
  \input{_limBar}%
\\
  & & Intelligence & 1 & % stacked bar chart for share of limitation categories
  \def\INTERNAL{100}%
  \def\EXTERNAL{0}%
  \def\CONSTRUCT{0}%
  \def\CONCLUSION{0}%
  \input{_limBar}%
 \\
  & & Background (generic) & 5 & % stacked bar chart for share of limitation categories
  \def\INTERNAL{80}%
  \def\EXTERNAL{20}%
  \def\CONSTRUCT{0}%
  \def\CONCLUSION{0}%
  \input{_limBar}%
 &                    & \multicolumn{2}{l}{\textbf{Operationalization and measures}} &  \\
  & \multicolumn{2}{l}{Individual knowledge} &  &  &                        & & Instrumentation & 39 & % stacked bar chart for share of limitation categories
  \def\INTERNAL{26}%
  \def\EXTERNAL{8}%
  \def\CONSTRUCT{56}%
  \def\CONCLUSION{10}%
  \input{_limBar}%
\\
  & & Ability & 3 & % stacked bar chart for share of limitation categories
  \def\INTERNAL{33}%
  \def\EXTERNAL{33}%
  \def\CONSTRUCT{0}%
  \def\CONCLUSION{33}%
  \input{_limBar}%
 &                                & & Mono-method bias & 1 & % stacked bar chart for share of limitation categories
  \def\INTERNAL{0}%
  \def\EXTERNAL{100}%
  \def\CONSTRUCT{0}%
  \def\CONCLUSION{0}%
  \input{_limBar}%
\\
  & & Domain Knowledge & 1 & % stacked bar chart for share of limitation categories
  \def\INTERNAL{100}%
  \def\EXTERNAL{0}%
  \def\CONSTRUCT{0}%
  \def\CONCLUSION{0}%
  \input{_limBar}%
  &                       & & Mono-operation bias & 4 & % stacked bar chart for share of limitation categories
  \def\INTERNAL{0}%
  \def\EXTERNAL{75}%
  \def\CONSTRUCT{25}%
  \def\CONCLUSION{0}%
  \input{_limBar}%
\\
  & & Education & 4 & % stacked bar chart for share of limitation categories
  \def\INTERNAL{75}%
  \def\EXTERNAL{25}%
  \def\CONSTRUCT{0}%
  \def\CONCLUSION{0}%
  \input{_limBar}%
 &                               & & Construct operationalization & 9 & % stacked bar chart for share of limitation categories
  \def\INTERNAL{44}%
  \def\EXTERNAL{0}%
  \def\CONSTRUCT{44}%
  \def\CONCLUSION{11}%
  \input{_limBar}%
\\
  & & \makecell{Familiarity with study \\object \& context} & 7 & % stacked bar chart for share of limitation categories
  \def\INTERNAL{43}%
  \def\EXTERNAL{43}%
  \def\CONSTRUCT{14}%
  \def\CONCLUSION{0}%
  \input{_limBar}%
 & \\
  & & Familiarity with tools & 3 & % stacked bar chart for share of limitation categories
  \def\INTERNAL{100}%
  \def\EXTERNAL{0}%
  \def\CONSTRUCT{0}%
  \def\CONCLUSION{0}%
  \input{_limBar}%
 &                  & \multicolumn{3}{l}{\textbf{Research design, procedure and conduct}} \\
  & & Programming experience & 23 & % stacked bar chart for share of limitation categories
  \def\INTERNAL{27}%
  \def\EXTERNAL{65}%
  \def\CONSTRUCT{4}%
  \def\CONCLUSION{4}%
  \input{_limBar}%
 &                 & & Learning effects & 12 & % stacked bar chart for share of limitation categories
  \def\INTERNAL{100}%
  \def\EXTERNAL{0}%
  \def\CONSTRUCT{0}%
  \def\CONCLUSION{0}%
  \input{_limBar}%
\\
  & & Reading time & 0 &  &                                                 & & Ordering & 1 & % stacked bar chart for share of limitation categories
  \def\INTERNAL{100}%
  \def\EXTERNAL{00}%
  \def\CONSTRUCT{0}%
  \def\CONCLUSION{0}%
  \input{_limBar}%
\\
  & \multicolumn{2}{l}{Individual circumstances} & & &                      & & Experimenter-subject interaction & 1 & % stacked bar chart for share of limitation categories
  \def\INTERNAL{0}%
  \def\EXTERNAL{0}%
  \def\CONSTRUCT{0}%
  \def\CONCLUSION{100}%
  \input{_limBar}%
\\
  & & Fatigue & 5 & % stacked bar chart for share of limitation categories
  \def\INTERNAL{100}%
  \def\EXTERNAL{0}%
  \def\CONSTRUCT{0}%
  \def\CONCLUSION{0}%
  \input{_limBar}%
 &                                 & & Evaluation apprehension & 3 & % stacked bar chart for share of limitation categories
  \def\INTERNAL{100}%
  \def\EXTERNAL{0}%
  \def\CONSTRUCT{0}%
  \def\CONCLUSION{0}%
  \input{_limBar}%
\\
  & & Motivation & 5 & % stacked bar chart for share of limitation categories
  \def\INTERNAL{80}%
  \def\EXTERNAL{20}%
  \def\CONSTRUCT{0}%
  \def\CONCLUSION{0}%
  \input{_limBar}%
 &                              & & Hawthorne effect & 3 & % stacked bar chart for share of limitation categories
  \def\INTERNAL{67}%
  \def\EXTERNAL{33}%
  \def\CONSTRUCT{0}%
  \def\CONCLUSION{0}%
  \input{_limBar}%
\\
  & & Treatment preference & 0 &  &                                         & & Process conformance & 5 & % stacked bar chart for share of limitation categories
  \def\INTERNAL{100}%
  \def\EXTERNAL{0}%
  \def\CONSTRUCT{0}%
  \def\CONCLUSION{0}%
  \input{_limBar}%
\\
  & & Stress & 1 & % stacked bar chart for share of limitation categories
  \def\INTERNAL{100}%
  \def\EXTERNAL{0}%
  \def\CONSTRUCT{0}%
  \def\CONCLUSION{0}%
  \input{_limBar}%
 &                                  & & Technical issues & 3 & % stacked bar chart for share of limitation categories
  \def\INTERNAL{100}%
  \def\EXTERNAL{0}%
  \def\CONSTRUCT{0}%
  \def\CONCLUSION{0}%
  \input{_limBar}%
\\
  & & Affective state & 1 & % stacked bar chart for share of limitation categories
  \def\INTERNAL{100}%
  \def\EXTERNAL{0}%
  \def\CONSTRUCT{0}%
  \def\CONCLUSION{0}%
  \input{_limBar}%
 &                         & & Causal model & 15 & % stacked bar chart for share of limitation categories
  \def\INTERNAL{60}%
  \def\EXTERNAL{20}%
  \def\CONSTRUCT{20}%
  \def\CONCLUSION{0}%
  \input{_limBar}%
\\
  & \multicolumn{2}{l}{Selection} & & &                                     & & Concentration impairment & 2 & % stacked bar chart for share of limitation categories
  \def\INTERNAL{100}%
  \def\EXTERNAL{0}%
  \def\CONSTRUCT{0}%
  \def\CONCLUSION{0}%
  \input{_limBar}%
\\
  & & Sampling strategy & 43 & % stacked bar chart for share of limitation categories
  \def\INTERNAL{21}%
  \def\EXTERNAL{72}%
  \def\CONSTRUCT{0}%
  \def\CONCLUSION{7}%
  \input{_limBar}%
 &                      & & Diffusion or imitation of treatments & 2 & % stacked bar chart for share of limitation categories
  \def\INTERNAL{100}%
  \def\EXTERNAL{0}%
  \def\CONSTRUCT{0}%
  \def\CONCLUSION{0}%
  \input{_limBar}%
\\
\multicolumn{3}{l}{} & \\                                                  
\multicolumn{3}{l}{\textbf{Materials}} & & &                                & \multicolumn{2}{l}{\textbf{Data analysis}} & \\
  & \multicolumn{2}{l}{Code snippet selection} & & &                        & & Data consistency & 4 & % stacked bar chart for share of limitation categories
  \def\INTERNAL{25}%
  \def\EXTERNAL{0}%
  \def\CONSTRUCT{25}%
  \def\CONCLUSION{50}%
  \input{_limBar}%
\\
  & & Content of study object & 35 & % stacked bar chart for share of limitation categories
  \def\INTERNAL{37}%
  \def\EXTERNAL{52}%
  \def\CONSTRUCT{11}%
  \def\CONCLUSION{0}%
  \input{_limBar}%
 &               & & Mortality and Study-object coverage & 2 & % stacked bar chart for share of limitation categories
  \def\INTERNAL{100}%
  \def\EXTERNAL{0}%
  \def\CONSTRUCT{0}%
  \def\CONCLUSION{0}%
  \input{_limBar}%
\\
  & & Programming language & 18 & % stacked bar chart for share of limitation categories
  \def\INTERNAL{44}%
  \def\EXTERNAL{56}%
  \def\CONSTRUCT{0}%
  \def\CONCLUSION{0}%
  \input{_limBar}%
 &                   & & Subjectivity in evaluating data & 16 & % stacked bar chart for share of limitation categories
  \def\INTERNAL{56}%
  \def\EXTERNAL{0}%
  \def\CONSTRUCT{38}%
  \def\CONCLUSION{6}%
  \input{_limBar}%
\\
  & & Layout of study object & 2 & % stacked bar chart for share of limitation categories
  \def\INTERNAL{50}%
  \def\EXTERNAL{0}%
  \def\CONSTRUCT{50}%
  \def\CONCLUSION{0}%
  \input{_limBar}%
 &                  & & ML data and model & 2 & % stacked bar chart for share of limitation categories
  \def\INTERNAL{50}%
  \def\EXTERNAL{0}%
  \def\CONSTRUCT{0}%
  \def\CONCLUSION{50}%
  \input{_limBar}%
\\
  & & Size of study object & 22 & % stacked bar chart for share of limitation categories
  \def\INTERNAL{9}%
  \def\EXTERNAL{91}%
  \def\CONSTRUCT{0}%
  \def\CONCLUSION{0}%
  \input{_limBar}%
 &                    & & Data quality & 12 & % stacked bar chart for share of limitation categories
  \def\INTERNAL{50}%
  \def\EXTERNAL{0}%
  \def\CONSTRUCT{42}%
  \def\CONCLUSION{0}%
  \input{_limBar}%
\\ % 8 other
  & & Number of study objects & 7 & % stacked bar chart for share of limitation categories
  \def\INTERNAL{14}%
  \def\EXTERNAL{86}%
  \def\CONSTRUCT{0}%
  \def\CONCLUSION{0}%
  \input{_limBar}%
 &                 & & Preliminarity of exploratory results & 2 & % stacked bar chart for share of limitation categories
  \def\INTERNAL{0}%
  \def\EXTERNAL{50}%
  \def\CONSTRUCT{0}%
  \def\CONCLUSION{50}%
  \input{_limBar}%
\\
  & & Snippet sampling strategy & 7 & % stacked bar chart for share of limitation categories
  \def\INTERNAL{0}%
  \def\EXTERNAL{86}%
  \def\CONSTRUCT{14}%
  \def\CONCLUSION{0}%
  \input{_limBar}%
 \\                                   
  & \multicolumn{2}{l}{Code environment} \\
  & & Code editor & 7 & % stacked bar chart for share of limitation categories
  \def\INTERNAL{14}%
  \def\EXTERNAL{72}%
  \def\CONSTRUCT{14}%
  \def\CONCLUSION{0}%
  \input{_limBar}%
 \\
  & & Snippet context & 2 & % stacked bar chart for share of limitation categories
  \def\INTERNAL{50}%
  \def\EXTERNAL{50}%
  \def\CONSTRUCT{0}%
  \def\CONCLUSION{0}%
  \input{_limBar}%
 \\
  & & Code environment (generic) & 1 & % stacked bar chart for share of limitation categories
  \def\INTERNAL{0}%
  \def\EXTERNAL{100}%
  \def\CONSTRUCT{0}%
  \def\CONCLUSION{0}%
  \input{_limBar}%
\\
  & & \makecell{Supporting materials and\\ resources} & 1 & % stacked bar chart for share of limitation categories
  \def\INTERNAL{0}%
  \def\EXTERNAL{100}%
  \def\CONSTRUCT{0}%
  \def\CONCLUSION{0}%
  \input{_limBar}%
\\
\bottomrule
\end{tabularx}
\end{table*}